\documentclass[aps,prd,notitlepage,nofootinbib,showpacs,superscriptaddress,floatfix]{revtex4-1}  
\usepackage{graphicx}  
\usepackage{dcolumn}   
\usepackage{bm}        
\usepackage{amssymb}   
\usepackage{amsmath}
\usepackage{comment}
\usepackage{graphicx}
\usepackage{subfig}
\usepackage{color}
\usepackage{slashed}
\usepackage{hyperref}

\newcommand{\Eq}[1]{Eq.(\ref{#1})}

\newcommand{\ti}[1]{\tilde{#1}}

\begin{document}

\author{Niklas Mueller}
\email{n.mueller@thphys.uni-heidelberg.de}
\affiliation{Institut f\"{u}r Theoretische Physik, Universit\"{a}t Heidelberg, Philosophenweg 16, 69120 Heidelberg, Germany}
\author{Raju Venugopalan}
\email{raju@bnl.gov}
\affiliation{Physics Department, Brookhaven National Laboratory, Bldg. 510A, Upton, NY 11973, USA}

\title{World-line construction of a covariant chiral kinetic theory}
\date{\today}
\begin{abstract}
We discuss a novel world-line framework for computations of the Chiral Magnetic Effect (CME) in ultrarelativistic heavy-ion collisions.
Starting from the fermion determinant in the QCD effective action, we show explicitly how its real part can be expressed as a supersymmetric world-line action of spinning, colored, Grassmanian particles in background fields.
Restricting ourselves for simplicity to spinning particles, we demonstrate how their constrained Hamiltonian dynamics arises for both massless and massive particles.
In a semi-classical limit, this gives rise to the covariant generalization of the Bargmann-Michel-Telegdi equation; the derivation of the corresponding Wong equations for colored particles is straightforward.
In a previous letter~\cite{Mueller:2017lzw}, we outlined how Berry's phase arises in a non-relativistic adiabatic limit for massive particles. We extend the discussion here to systems with a finite chemical potential. 
We discuss a path integral formulation of the relative phase in the fermion determinant that places it on the same footing as the real part. We construct the corresponding anomalous world-line axial vector current and show in detail how the chiral anomaly appears. Our work provides a systematic framework for a relativistic kinetic theory of chiral fermions in the fluctuating topological backgrounds that generate the CME in a deconfined quark-gluon plasma. We outline some further applications of this framework in many-body systems.

\end{abstract}
\maketitle
%
%
\section{Introduction}
The prospect of experimental access in ultrarelativistic heavy-ion collisions to emergent CP- and P-odd phenomena in Quantum Chromodynamics (QCD) has inspired much interest. 
Some of this interest derives from the fact that topology changing sphaleron transitions~\cite{Klinkhamer:1984di,Dashen:1974ck,Soni:1980ps,Boguta:1983xs,Forgacs:1983yu}, associated with  the quantum anomalies generating such 
phenomena, are a central ingredient in models of electroweak baryogenesis in the evolution of the early universe~\cite{Sakharov:1967dj,Riotto:1999yt,Cohen:1993nk,Rubakov:1996vz}.
Quantum anomalies are also conjectured to play an important role in the electronic properties of strongly correlated condensed matter systems~\cite{cond-matter_review}. 
In general, the real-time dynamics involving the effects of anomalies are an excellent probe of the topological structure of gauge theories. In the context of heavy-ion collisions, a major advance is the conjectured existence of a Chiral Magnetic Effect (CME). The CME here corresponds to fluctuations of axial charge imbalances in the strongly correlated quark-gluon plasma (QGP) that generate electric currents in the direction of the Abelian magnetic fields that exist in off-central collisions of the heavy nuclei~\cite{Kharzeev:2007jp,Fukushima:2008xe,Kharzeev:2015znc}. 

The CME has already been observed in condensed matter experiments \cite{Li:2014bha}.  Observing its effects in heavy-ion collisions however poses a significant challenge~\cite{Kharzeev:2015znc,Skokov:2016yrj}.
It requires an understanding of the earliest times in the heavy-ion collision, as the Abelian magnetic fields generated by ``spectator" nucleons decrease very rapidly in time \cite{Skokov:2009qp,Deng:2012pc}.
Weak coupling frameworks applicable at high energies indicate that, at these early times,
the strongly correlated quark and gluon matter is far off-equilibrium in a highly overoccupied ``Glasma" state, which subsequently thermalizes to a quark-gluon plasma (QGP). 
Recent studies suggest that sphaleron transitions  are far more frequent in the Glasma \cite{Mace:2016svc}, than in the QGP \cite{Moore:2010jd}.
Classical-statistical real-time simulations that include the dynamics of chiral fermions \cite{Tanji:2016dka} clearly demonstrate the emergence of the CME in background magnetic fields~\cite{Mueller:2016ven,Mace:2016shq}.

However this real-time description of the Glasma breaks down when, due to the spacetime expansion of the Glasma, typical occupation numbers become of order unity. 
In this dilute regime of the Glasma, classical-statistical methods must be matched to kinetic descriptions that describe the dynamics of the system as a weakly interacting gas of quasi-particles. 
Real-time simulations studying the thermalization process in the Glasma \cite{Berges:2013eia} show that the classical-statistical description matches smoothly on to an effective kinetic theory~\cite{Kurkela:2015qoa},
which in turn can be matched to relativistic viscous hydrodynamics at later times. This description, when extrapolated to realistic values of coupling, gives values for thermalization times that are compatible with hydrodynamic descriptions 
of heavy-ion data. Phenomenological studies in such a hybrid framework have now been extended to photon production, whose yields are sensitive to all spacetime stages of a heavy-ion collision~\cite{Berges:2017eom}. 

Similar considerations apply to the classical-statistical description of the spacetime evolution of the chiral magnetic current through the Glasma. The development of a chiral kinetic theory that interpolates between 
classical-statistical Glasma dynamics of axial charges at early times and hydrodynamic descriptions of such dynamics in the QGP~\cite{Son:2009tf,Gursoy:2014aka,Hongo:2013cqa,Hirono:2014oda,
Yin:2015fca} at late times is therefore essential for systematic phenomenological analysis of the CME in heavy-ion collisions. There has been a significant amount of work in developing such a chiral kinetic theory both in
the context of condensed matter systems and for a deconfined QGP~\cite{Son:2012wh,Stephanov:2012ki,Son:2012zy,Chen:2013iga,Chen:2014cla,Stone:2013sga,Dwivedi:2013dea,Manuel:2014dza,Stone:2014fja,Manuel:2015zpa,Chen:2015gta,Sun:2016nig,Hidaka:2016yjf}.
In several of the treatments, systems with large chemical potential are considered. The dynamics includes a Berry term corresponding to the Berry phase~\cite{Berry:1984jv}
that arises in such systems in an adiabatic limit, valid for excitations near the Fermi surface. While such treatments may be appropriate for systems containing large chemical potentials, they are problematic in relativistic contexts
such as heavy-ion collisions where the assumptions of adiabaticity may not apply and where chemical potentials are not a priori large. 

A further concern with chiral kinetic treatments is the possible conflation of topological effects due to the chiral anomaly and those arising from geometric phases in adiabatic and non-relativistic limits. Unlike the latter, the topological effects due to anomalies are generic and independent of kinematic limits. The connection between anomalies and Berry's phase, which has been made frequently in the literature~\cite{Stone:1985av,Aitchison:1986qn,Nelson:1984gu} (see \cite{Shapere:1989kp} for a review), is the subject of a critical series of papers by Fujikawa and collaborators~\cite{Deguchi:2005pc,Fujikawa:2005tv,Fujikawa:2005cn}, where they point to distinctions between the topology of Berry's phase and those of the anomaly~\cite{Wess:1971yu,Witten:1983tw}). 

In this work, we will develop a novel framework towards constructing a consistent Lorentz-covariant chiral kinetic theory that is general valid in relativistic contexts and makes no requirement that the dynamics be adiabatic. To achieve this goal, we will adopt the world-line approach\footnote{The original ideas can be traced all the way back to seminal works by Feynman~\cite{Feynman:1950ir} and Schwinger~\cite{Schwinger:1951nm}.} to quantum field theory~\cite{Strassler:1992zr,Mondragon:1995va,Mondragon:1995ab,JalilianMarian:1999xt,Schubert:2001he,Bastianelli:2006rx,Hernandez:2008db,Corradini:2015tik}. 
This world-line framework is closely connected to the Polakov path integral in string theory \cite{Polyakov:1987ez}. These connections were very effectively exploited in the work of Bern, Dixon, Dunbar and Kosower \cite{Bern:1991aq,Bern:1994zx} relating string amplitudes to multi-leg Feynman diagrams in QCD. More to point, it is employed in the seminal work on quantum anomalies by Alvarez-Gaume and Witten~\cite{AlvarezGaume:1983ig,AlvarezGaume:1983at} where it is shown how anomalies arise in the framework from the 
phase of the fermion determinant -- as anticipated in the work of Fujikawa~\cite{Fujikawa:1979ay,Fujikawa:1980eg}. 

In a previous letter~\cite{Mueller:2017lzw}, we showed that a particular world-line construction\footnote{See also related work in \cite{Mondragon:1995ab,Hernandez:2008db}.} of D'Hoker and Gagn\'e \cite{D'Hoker:1995ax, D'Hoker:1995bj} is well suited to the construction of a chiral kinetic theory. We sketched there how the coherent state formalism of  D'Hoker and Gagn\'e gives rise to the Bargmann-Michel-Telegdi equations for spinning point particles in external gauge field backgrounds and Wong's equations for their colored counterparts. We further outlined how for massive particles the corresponding Hamiltonian description generates a Berry phase when an adiabaticity condition is imposed. 
The principal value of the D'Hoker-Gagn\'e world-line construction is in its treatment of the relative phase in the fermion determinant which, as noted, is responsible for the chiral anomaly. By an ingenious trick, this phase can be rewritten as a path integral, with a point particle ``action". This action has an identical structure to the action arising from the real part of the fermion determinant, with the only (and critically important) change being that the gauge fields are multiplied by a regulating parameter which breaks chiral symmetry explicitly. In the letter, we briefly outlined how the chiral anomaly arises in the D'Hoker-Gagn\'e construction. 

We will here develop many of the ideas outlined in \cite{Mueller:2017lzw} and provide an explicit derivation, adapted to our QED/QCD framework, of the D'Hoker-Gagn\'e formalism. For the real part of the effective action, we explicitly write down the point particle action, and demonstrate that the equations of motion 
for QED are the covariant generalization of the Bargmann-Michel-Telegdi~\cite{Bargmann:1959gz} equations for spinning particles in external fields. (For colored particles in QCD, the counterparts are the Wong~\cite{Wong:1970fu} equations.) In particular, we will discuss the constrained Hamiltonian dynamics of spinning particles~\cite{Berezin:1976eg,Balachandran:1976ya,Balachandran:1977ub,Barducci:1982yw,Brink:1976uf} in the world-line approach. 
This discussion is of considerable importance in deriving the non-relativistic limit for spinning particles. As noted in our letter, this approach leads in a clean logical chain to Berry's phase after further assumptions of adiabaticity. We will here extend the latter discussion to the case of systems with large chemical potentials.

Another novel feature of this manuscript is an explicit derivation of the chiral anomaly in the  D'Hoker-Gagn\'e world-line construction. In their work, they used a perturbative expansion to show how a Wess-Zumino-Witten term arises~\cite{Wess:1971yu,Witten:1983tw} from the relative phase in the fermion determinant. In our work, in addition to clarifying some subtle points in the D'Hoker-Gagn\'e construction,
we will instead employ a non-perturbative variational method to derive the anomaly equation explicitly as the scalar product of electric and magnetic fields. The corresponding world-line anomalous axial current has a structure we will find useful in constructing a chiral kinetic theory. 

This observation provides the segue to note that world-line treatment of the real and imaginary terms in the effective action for the fermion determinant both provide essential ingredients in a kinetic description of relativistic fermions in the background of Abelian or non-Abelian gauge fields. The quasi-particle limit of the theory, and furthermore
the Liouville description of phase space, is contained entirely in the \textit{real} part of the fermion effective action (continued to Minkowskian metric), independently of the anomaly. The 
Hamilton evolution of the corresponding equations of motion, formulated in proper time $\tau$, allow for a Lorentz-covariant kinetic theory. Spin effects related to the definition of a Lorentz frame, 
such as recently proposed ``side-jumps" are natural outcomes of a covariant description of spinning particles~\cite{Chen:2014cla}.  We showed in our letter \cite{Mueller:2017lzw} that for a non-relativistic limit corresponding to massive particles, adiabaticity conditions on the Larmor interaction energy, generate a Berry phase. Since this derivation only involves the real part of the fermion determinant, and the chiral anomaly arises from its imaginary piece, our work is an explicit demonstration of the prior observation by Fujikawa and collaborators~\cite{Fujikawa:1979ay,Fujikawa:1980eg} regarding the distinction between the topological effects arising from each. For massless relativistic particles, and for situations where the Larmor energy is large, the topology of the anomaly alone is relevant. 

An exception is the case of systems with large chemical potential, the original focus of the kinetic theory construction in \cite{Son:2012wh,Son:2012zy}. We will extend our discussion of non-relativistic limits in \cite{Mueller:2017lzw} to this case. We will show explicitly how the adiabaticity condition for the Larmor energy arises in this case. However even though there is a Berry phase in such situations, it is still distinct from the effects from the anomaly. Our work provides a first principles framework to address the fascinating interplay of these distinct effects. As noted in our letter, the real-time formulation~\cite{Mathur:1993tp}  of a semi-classical world-line kinetic theory for spinless colored particles results in the non-Abelian Boltzmann-Langevin B\"{o}deker kinetic theory of hot QCD~\cite{Bodeker:1998hm,Bodeker:1999ey,Litim:1999id,JalilianMarian:1999xt} including both noise and collision terms. In work in progress\footnote{For another attempt, we refer the reader to \cite{Akamatsu:2014yza}.}, the formalism discussed here will be employed to derive the analogous ``anomalous" B\"{o}deker theory~\cite{follow-up}. The resulting generalization of chiral kinetic theory can then be matched to results from classical-statistical simulations at early times and to anomalous hydrodynamics at late times. 

The outline of this manuscript is as follows: In section \ref{sec:WLframework}, we begin by giving an introduction to the world-line method and we work out its formulation for a Dirac fermion coupled to both vector and 
axial-vector gauge fields. In particular, we introduce a 16 dimensional matrix formulation of the fermion effective action. As we shall discuss, this formulation is convenient for implementing a coherent state 
formalism for spinning and colored fields. We will show how the real part of the effective action is expressed in terms of a  Grassmanian path integral over a supersymmetric point particle action for such fields.  
We next discuss the D'Hoker-Gagn\'e path integral construction for the imaginary phase in the fermion determinant and show that it has a similar structure to the path integral formulation of the real part of the effective action. We use this construction to derive expressions for the vector and  anomalous axial vector current which  fullfil
the (anomalous) Ward-identities known from second quantization.  We pay particular attention to the anomaly equation, which has novel features, and provide a detailed derivation to expose these features.   In section \ref{sec:KineticTheory} we perform a saddle point expansion to  obtain the pseudo-classical dynamics of spinning particles. These were studied extensively previously in the literature and we connect our results to this body of work~\cite{Berezin:1976eg,Balachandran:1976ya,Balachandran:1977ub,
Barducci:1982yw,Brink:1976uf}  in sec. \ref{sec:pseudoclassicallimit}. We note some parallels between our work and those of Stone and collaborators~\cite{Stone:2013sga,Dwivedi:2013dea,Stone:2014fja} though the derivations are different and employ different techniques. 

The pseudo-classical limit of the world-line effective action leads to a Lorentz covariant form of the ``anomalous"  equations of motions put forward by \cite{Son:2012wh,Stephanov:2012ki,
Son:2012zy,Chen:2013iga,Chen:2014cla,Chen:2015gta}, when taking the non-relativistic and adiabatic limit in section~\ref{sec:NRadiabatic}. 
A kinetic theory can be constructed from the world-line framework for half-integer particles;  as noted, the equations of motion obtained from the stationarity condition
of the world-line path integral constitute characteristic equations for Liouville evolution of the phase space density. We investigate the case of massless particles in the presence of large chemical potential and discuss the corresponding non-relativistic adiabatic limit.

Our findings are supplemented by several appendices: In appendix \ref{sec:app:anomalydetails},
we provide details of a derivation that is not discussed in the main text. In appendix \ref{app:susy}, we discuss the symmetry properties of the world-line action for spin-$1/2$ particles, corresponding to an $N=1$ supersymmetric quantum mechanics. As our derivations generalize to arbitrary internal symmetry groups, we given an introduction in appendix \ref{app:internalsymmetries} to how color degrees of freedom
can be treated semi-classically using Grassmanian variables. In appendix \ref{app:anomalies}, we discuss the difference between covariant and consistent anomalies. Finally in appendix \ref{sec:saddlepointdef},  we discuss in detail the meaning of the pseudo-classical limit in the world-line framework, which is tied to a hidden gauge symmetry and to constraints, which arise upon quantization.
%
%
%
\section{The world-line framework}\label{sec:WLframework}
\subsection{Introduction}\label{sec:WLframework:intro}
In this section, we shall derive in the world-line formalism, the one-loop effective action for a Dirac fermion coupling to vector and axial-vector gauge fields. We will show that in Euclidean metric the axial anomaly can be understood
as arising from the \textit{imaginary} part of the effective action \cite{AlvarezGaume:1983ig,AlvarezGaume:1983at}. This result is transparently  related to the violation of chiral symmetry. 
 We begin by introducing the main ideas of the relevant world-line framework. Some parts of our derivation parallel the work
of D'Hoker and Gagn\'e \cite{D'Hoker:1995ax,D'Hoker:1995bj}. We will however place special emphasis on some of the details in the definition of single particle path integrals. The careful treatment of these is relevant for the realization of the
axial anomaly. The expression for the fermionic part of the action in the background of 
vector ($A$) and axial-vector ($B$) fields is 
\begin{align}
S[A,B]=\int d^4x\;\bar{\psi}\left(i\slashed{\partial}+\slashed{A}+\gamma_5\slashed{B} \right)\psi\,,\label{eq:classicalaction}
\end{align}
where we allow the fermion fields to carry any internal (gauge) symmetry. We introduced here an auxilliary  Abelian axial-vector field $B$ because we are interested in the color singlet axial anomaly. We will treat $B$ as a variational parameter which we will set to zero eventually.
In the following, we have absorbed all couplings into the definition of the fields for convenience and they can be easily restored when necessary. 

The fermionic part of the full path integral 
containing the action in \Eq{eq:classicalaction} is a Grassmanian Gaussian integral and can be performed. This gives the determinant of the bilinear operator, $\det (i\slashed{\partial}+{\slashed{A}}+\gamma_5{\slashed{B}})$, 
from which the fermion effective action can be defined,
\begin{align}
-W[A,B]=\log\det(\theta),\qquad\qquad\theta\equiv i\slashed{\partial}+{\slashed{A}}+\gamma_5{\slashed{B}}\,.\label{eq:effectiveaction}
\end{align}
We can now split \Eq{eq:effectiveaction} in a real and imaginary part,
\begin{align}
W[A,B]=W_\mathbb{R}[A,B]+iW_\mathbb{I}[A,B],\label{eq:realimagpartseff}
\end{align}
which we discuss in detail below. We will continue with massless quarks;  the extension to massive particles is straightforward and for the problems of interest will be discussed explicitly later. Since the imaginary piece above may be unfamiliar to some, we mention for future reference that,
albeit in the physical case one has $W[A,B=0]_\mathbb{I}=0$, the \textit{variation} $\delta W[A,B]_\mathbb{I}/\delta B_\mu$ is non-zero even if $B=0$. This variation defines 
the anomalous axial-vector current. For the sake of illustration, our final results will be given for the QED anomaly, but we will discuss how our findings can be generalized to non-Abelian theories as well. In Appendix \ref{sec:app:anomalydetails} we provide
supplementary material and elaborate on some intermediate steps in the calculation.
%
%
\subsection{Real Part}\label{sec:WLframework:real}
In this section, we will derive an expression for the real part of the fermion determinant, defined in \Eq{eq:effectiveaction} and \Eq{eq:realimagpartseff}.
The real part is related to the modulus of the operator $\theta$ and can be expressed as
\begin{align}
W_\mathbb{R}=-\frac{1}{2}\log\det\left(\theta^\dagger\theta\right)=-\frac{1}{2}\text{Tr}\log\left(\theta^\dagger\theta\right)\,.\label{eq:abstrace}
\end{align}
The main idea behind the world-line technique is to find an integral representation for the logarithm of the positive definite operator $\theta^\dagger \theta$. As we will shown below, this is equivalent
to defining a quantum mechanical path integral for a relativistic particle on a closed loop, which is the world-line. 
We will require a basis of states for the trace in \Eq{eq:abstrace}, which is over
an infinite-dimensional space and contains both spacetime as well as internal indices. For spinors, this basis is related to the Clifford algebra of fermions, but the basis can include possible further internal symmetry groups such as color.

The spacetime trace can be turned into a quantum mechanical path integral for the bosonic coordinates of a point particle, as was shown in \cite{Strassler:1992zr}.  The trace over the Dirac matrix structure of spinors leads to path integrals using a Grassmanian coherent state formalism. Such a coherent state formalism  is discussed in \cite{D'Hoker:1995ax,D'Hoker:1995bj} and requires an artificial enlargment of the dimension of the space, in which the Dirac matrix structure is embedded, from $4\times 4$ to $8\times 8$. Interpreting $\theta^\dagger\theta$ as an eight dimensional matrix and making a similarity transformation as outlined in detail in \cite{D'Hoker:1995bj}, the fermionic effective action can be written as
\begin{align}
W_\mathbb{R}=-\frac{1}{8}\log\det(\ti{\Sigma}^2)=-\frac{1}{8}\text{Tr}\log(\ti{\Sigma}^2),\label{eq:effactREAL}
\end{align}
where $\ti{\Sigma}^2$ is given by
\begin{align}
\ti{\Sigma}^2=(p-\mathcal{A})^2\;\mathbb{I}_8+\frac{i}{2}\Gamma_\mu\Gamma_\nu F_{\mu\nu}[\mathcal{A}],\label{eq:Sigma}\,.
\end{align}%
Here we have artificially enlarged the representation space of the gauge field to include the left and right handed chiral fields,
\begin{align}
\mathcal{A}=\begin{pmatrix}
{A}+{B} & 0\\
0 & {A}-{B}
\end{pmatrix}\,,\label{eq:internalfieldrep}
\end{align}
whereby $\ti{\Sigma}^2$ is a sixteen dimensional ($8\times 2$) matrix. The six $8\times 8$ dimensional gamma matrices $\Gamma_a$ are defined as
\begin{align}\label{eq:gammamatrices}
\Gamma_\mu=
\begin{pmatrix}
0 & \gamma_\mu \\
\gamma_\mu  & 0 
\end{pmatrix},\quad
\Gamma_5=
\begin{pmatrix}
0 & \gamma_5 \\
\gamma_5  & 0 
\end{pmatrix},\quad
\Gamma_6=
\begin{pmatrix}
0 & i \mathbb{I}_4 \\
i\mathbb{I}_4   & 0 
\end{pmatrix},
\end{align}
with an additional matrix $\Gamma_7$, anti-commuting with all other elements of the algebra, 
\begin{align}\label{eq:gammamatrix7}
\Gamma_7=-i\prod\limits_{A=1}^6\Gamma_A=
\begin{pmatrix}
\mathbb{I}_4 & 0 \\
0   & -\mathbb{I}_4 
\end{pmatrix},\qquad\qquad \{\Gamma_7,\Gamma_A\}=0.
\end{align}
Here $\gamma_\mu$ and $\gamma_5$ are the usual Dirac matrices. 

This artificial enlargemment of both the dimensions of the Dirac matrices as well as the representation of gauge fields may seem unmotivated.
Indeed the splitting of \Eq{eq:internalfieldrep} is strictly speaking not necessary at all, but simplifies our calculations significantly. The dimensional extension of the Dirac matrices, on the other hand, as defined in \Eq{eq:gammamatrices} is a necessity. The elementary idea behind the world-line approach is to express traces, such as those given in \Eq{eq:effactREAL}, in terms of  quantum mechanical single particle
states. As observed in \cite{D'Hoker:1995ax,Mehta:1986mi}, this is not possible for four-dimensional Dirac matrices; a set of coherent fermion states, representing the 
corresponding Clifford algebra, exists however for the extension given in \Eq{eq:gammamatrices}.

With this path integral formulation in mind, we will adopt Schwinger's integral representation to write \Eq{eq:effactREAL} as 
\begin{align}
W_\mathbb{R}=\frac{1}{8}\int\limits_0^\infty\frac{dT}{T}\;\text{Tr}_{16}\;e^{-\frac{\mathcal{E}}{2}T\tilde{\Sigma}^2}\label{eq:tr},
\end{align}
where, by means of the $T-$integral, we have introduced what is commonly known as a closed world-line of length $T$. While \Eq{eq:tr} can be taken as the definition of the world-line, its structure will be discussed in more detail below. We introduced here an arbitrary positive real number $\mathcal{E}$ called the einbein. As is well known, and as we shall discuss explicitly in section \ref{sec:KineticTheory} and in appendix \ref{app:susy},
$\mathcal{E}$ is not a physical quantity but rather a gauge parameter related to reparametrization invariance on the world-line.

The trace in \Eq{eq:tr} includes the internal (Dirac-)space and it can be evaulated using  a coherent state basis that realizes the Clifford algebra of
Dirac fermions.  More specifically, the spin part of the trace in \Eq{eq:tr} is turned into a path integral over Grassman variables \cite{D'Hoker:1995ax,D'Hoker:1995bj}, employing the methods developed first by Berezin 
and Marinov \cite{Berezin:1976eg}. Towards this end, we introduce the fermion creation and annihilation operators ($a^\pm_r, r=1,2,3$),
\begin{align}\label{eq:coherent1}
a^\pm_r=\frac{1}{2}(\Gamma_r\pm i\Gamma_{r+3}),\qquad
\{a^+_r,a^-_s  \}=\delta_{rs},\qquad\{a^+_r,a^+_s  \}=\{a^-_r,a^-_s  \}=0.
\end{align}
These operators $a^\pm_r$  span the space of the Clifford algebra satisfied by the $\Gamma$ matrices. They define the coherent states $|\theta\rangle, |\bar{\theta}\rangle$ which satisfy
\begin{align}\label{eq:coherent2}
\langle \theta | a_r^-=\langle \theta |\theta_r\qquad a_r^-| \theta\rangle=\theta_r|\theta\rangle\qquad
\langle \bar{\theta}| a_r^+ =\langle \bar{\theta}| \bar{\theta}_r\qquad a_r^+|\bar{\theta}\rangle=\bar{\theta}_r|\bar{\theta}\rangle\,,
\end{align}
with the matrix elements between coherent states defined to be 
\begin{align}
\langle \theta | \bar{\theta}\rangle=e^{\theta_r\bar{\theta}_r},\qquad
\langle \bar{\theta} | {\theta}\rangle=e^{\bar{\theta}_r{\theta}_r} \,.
\end{align}
These satisfy the completeness relations
\begin{align}
\int |\theta \rangle\langle\theta | \;d^3\theta=\int d^3\bar{\theta} \,|\bar{\theta}\rangle \langle\bar{\theta}|=\mathbb{I}\,.
\end{align}
Note that while $\theta_r,\bar{\theta}_r,d\theta_r,d\bar{\theta}_r$ anticommute with $\langle \theta |$, $|\bar{\theta}\rangle$, they commute with $|\theta \rangle$, $\langle\bar{\theta}|$. 
All states and variables commute with the vacuum. With these definitions, traces in the coherent state basis can be defined. 

The trace over a generic operator
has the form 
\begin{align}\label{eq:grassmann:trace}
\text{Tr}(O)=\int d^3\theta \langle -\theta| O| \theta\rangle\,.
\end{align} 
This expression for the trace is discussed at length in \cite{Ohnuki:1978jv}. The negative sign in \Eq{eq:grassmann:trace} arises from transforming the coherent state basis to a Fock state basis. As this includes anti-commuting variables, the minus sign in \Eq{eq:grassmann:trace}  can be interpreted as enforcing anti-periodic boundary conditions for the Grassmann variables on the closed world-line. We can therefore write the trace in \Eq{eq:tr} as 
\begin{align}\label{eq:tr1}
\text{Tr}_{16}\;e^{-\frac{\mathcal{E}}{2}T\tilde{\Sigma}^2}=\text{tr}\int d^4z\; d^3\theta \;\langle z, -\theta | e^{-\frac{\mathcal{E}}{2}T\tilde{\Sigma}^2} | z. \theta\rangle\,.
\end{align}
The remaining trace ($\text{tr}$) on the r.h.s now contains only the trace over the representation space \Eq{eq:internalfieldrep} and other internal symmetries such as color. If we proceed with Abelian gauge fields alone, $\text{tr}$ is only over the two dimensional representation space \Eq{eq:internalfieldrep} and is in fact trivial -- it amounts to a simple sum over the two chiral configurations, as we will see below. Non-Abelian gauge fields can be included straigthforwardly. 
as we show in Appendix \ref{app:internalsymmetries}. For simplicity, we will discuss only Abelian gauge fields for the rest of the manuscript; the extension to QCD will be discussed in follow-up papers. 

We will now express the matrix element on the r.h.s of \Eq{eq:tr1} as a path integral\footnote{This strategy highlights the fact that in the world-line approach, contrary to the conventional approaches in quantum field theory, spin is not accounted for by means of a multidimensional wave function (as it is done for fermion spinors)  but instead as an independent degree of freedom in the path integral.}, with $T$ playing the role of ``time" and the ``Hamiltonian" represented by 
$\mathcal{E}\tilde{\Sigma}^2/2$ \cite{D'Hoker:1995ax,D'Hoker:1995bj}.  Our derivation, for this real part of the effective action, uses the conventional time-slicing procedure to construct  the path integral. 
Splitting the time interval into $N$ discrete steps of length $\Delta\equiv T/N$ (the continuum limit defined as $N\rightarrow \infty$ and 
$\Delta\rightarrow 0$), we define the average position between two time-slices 
\begin{align}
\bar{x}^k_\mu=\frac{x^k_\mu+x^{k-1}_\mu}{2},
\end{align}
and for later use combine the three complex Grassman variables $\theta,\bar{\theta}$ into six real ones,
\begin{align}\label{eq:psitheta}
\psi_a^k&=\frac{1}{\sqrt{2}}(\theta^k_a+\bar{\theta}^k_a)&a=1,2,3\nonumber\\
\psi_a^k&=\frac{i}{\sqrt{2}}(\theta^k_{a-3}-\bar{\theta}^k_{a-3}) &a=4,5,6 \, .
\end{align}
Further, with these definitions, matrix elements containing Gamma-matrices $\Gamma$ are evaluated by making use of \Eq{eq:coherent1} and \Eq{eq:coherent2} to read 
\begin{align}\label{eq:Grassmantrace}
\langle \theta^k|\Gamma_a \Gamma_b | \theta^{k-1}\rangle = -\int d\bar{\theta}^k\langle\theta^k|\bar{\theta}^k\rangle \langle \bar{\theta}^k | \theta^{k-1} \rangle\, 2(\psi_a^k \psi_b^{k-1})
=-\int d\bar{\theta}^k e^{\theta^k_r\bar{\theta}^k_r+\bar{\theta}^k_r\theta^{k-1}_r}\,2( \psi_a^k \psi_b^{k-1}).
\end{align}
\Eq{eq:Grassmantrace} can be generalized to higher matrix products using the simple mnemonic $\Gamma_a\rightarrow\sqrt{2}\,\psi_a$. 

After these preliminaries, inserting complete sets of coherent states, we obtain, 
\begin{align}\label{eq:pathintreal}
\text{Tr}\Big\{e^{-\frac{\mathcal{E}}{2}T\tilde{\Sigma}^2}&\Big\}=
-\text{tr}\int \left(\prod\limits_{l=0}^{N-1}d^4 x_l\right)\left(\prod\limits_{l=1}^{N}\frac{d^4p_l}{(2\pi)^4}\right)
\left(\prod\limits_{l=0}^{N-1}d^3 \theta_l\right)\left(\prod\limits_{l=1}^{N}d^3 \bar{\theta}_l\right)\nonumber\\&\times
\exp \Big\{-\Delta\sum\limits_{k=1}^{N}\big[ 
-ip_\mu^k\frac{(x^k_\mu-x^{k-1}_\mu)}{\Delta}+\frac{\mathcal{E}}{2}\left(p_\mu^k-\mathcal{A}_\mu[\bar{x}^k] \right)^2
\quad-\frac{(\theta^k_r-\theta^{k-1}_r)}{\Delta}\bar{\theta}^k_r   +\frac{i\mathcal{E}}{2}\psi^k_\mu
F_{\mu\nu}[\bar{x}^k]\psi_\nu^{k-1}
\big]\Big\}\nonumber\\
&=-\text{tr}\int \left(\prod\limits_{l=0}^{N-1}d^4 x_l\right)\left(\prod\limits_{l=1}^{N}\frac{d^4p_l}{(2\pi)^4}\right)
\left(\prod\limits_{l=0}^{N-1}d^3 \theta_l\right)\left(\prod\limits_{l=1}^{N}d^3 \bar{\theta}_l\right)\nonumber\\&\times
\exp \Big\{-\Delta\sum\limits_{k=1}^{N}\big[ 
-ip_\mu^k\frac{(x^k_\mu-x^{k-1}_\mu)}{\Delta}+\frac{\mathcal{E}}{2}\left(p_\mu^k-\mathcal{A}_\mu[\bar{x}^k] \right)^2
\quad+\frac{1}{2}\psi_a^k\frac{(\psi_a^k-\psi_a^{k-1})}{\Delta}+\frac{i\mathcal{E}}{2}\psi^k_\mu
F_{\mu\nu}[\bar{x}^k]\psi_\nu^{k-1}
\big]\Big\}\nonumber\\
&\equiv\mathcal{N}\int_P\mathcal{D}x \int_{AP}\mathcal{D}\psi\;\text{tr}\;\exp{\Big\{-\int_0^T d\tau\;\mathcal{L}(\tau)\Big\}}.
\end{align}
In obtaining the second equality,  we symmetrized the ``kinetic term" with respect to the variables $\theta,\bar{\theta}$ in order to replace the complex variables $\theta$ with $\psi_a^k$, using \Eq{eq:psitheta}, before taking the continuum limit of the path integral~\cite{D'Hoker:1995ax,D'Hoker:1995bj}. Further, in the last step, we completed the squares and shifted the $p$ integration\footnote{This standard trick replaces 
$p_\mu^k\rightarrow p_\mu^k-A_\mu[\bar{x}^k]-i(x^k_\mu-x^{k-1}_\mu)/\mathcal{E}\Delta$.}.
Periodic boundary conditions $P$ for bosonic variables and anti-periodic boundary conditions $AP$ for fermion observables are imposed respectively by identifying $x^0=x^N$ and $\psi^0=-\psi^N$. Expressing the Grasssmanian integration measure by the six-dimensional variables $\mathcal{D}\psi=\mathcal{D}\psi_\mu\mathcal{D}\psi_5\mathcal{D}\psi_6$, generates a trivial Jacobian, which can be absorbed in the normalization. 

The real part of the effective action can thus be expressed in path integral form as 
\begin{align}
W_\mathbb{R}=\frac{1}{8}\int\limits_0^\infty\frac{dT}{T}\mathcal{N}\int\limits_{P}
\mathcal{D}x\int\limits_{AP}\mathcal{D}\psi\;\text{tr}\exp{\Big\{-\int\limits_0^Td\tau\;\mathcal{L}(\tau)\Big\}}\label{eq:realpart}.
\end{align}
with the point particle ``quantum mechanical" world-line Lagrangian
\begin{align}\label{eq:realpartLag}
\mathcal{L}(\tau)=\frac{\dot{x}^2}{2\mathcal{E}}+\frac{1}{2}\psi_a\dot{\psi}_a-i\dot{x}_\mu\mathcal{A}_\mu+\frac{i\mathcal{E}}{2}\psi_\mu F_{\mu\nu}[\mathcal{A}]\psi_\nu \,,
\end{align}
where 
\begin{align}\label{eq:lagrangianexplicit}
\mathcal{L}(\tau)=\begin{pmatrix}
\mathcal{L}_{L} & 0\\
0 & \mathcal{L}_{R}
\end{pmatrix},\qquad
\mathcal{L}_{L/R}(\tau)=\frac{\dot{x}^2}{2\mathcal{E}}+\frac{1}{2}\psi_a\dot{\psi}_a-i\dot{x}_\mu(A\pm B)_\mu+\frac{i\mathcal{E}}{2}\psi_\mu F_{\mu\nu}[A\pm B]\psi_\nu 
\, ,
\end{align}
carries the two-dimensional matrix structure of the helicity representation of the gauge fields \Eq{eq:internalfieldrep} and can be trivially split into separate Lagrangians for both chiralities/helicities. The path integral can further be written more explicitly as 
\begin{align}\label{eq:realpartpath}
W_\mathbb{R}=\frac{1}{8}\int\limits_0^\infty\frac{dT}{T}\mathcal{N}\int\limits_P\mathcal{D}x\int
\limits_{AP}\mathcal{D}\psi\;\left[ \exp{\Bigg\{-\int\limits_0^Td\tau\;\mathcal{L}_{L}(\tau)\Bigg\}}+\exp{\Bigg\{-\int\limits_0^Td\tau\;\mathcal{L}_{R}(\tau)\Bigg\}}\right].
\end{align}
For a vector gauge theory, where $B=0$, $\mathcal{L}_{L}=\mathcal{L}_{R}$, as both left and right handed massless particles couple to vector fields identically.
In this case, the the trace in \Eq{eq:realpart} just gives an overall factor of two. For the reasons outlined previously,  we will keep $B\neq 0$.
The normalization in \Eq{eq:realpart} is 
\begin{align}\label{eq:normalization}
\mathcal{N}\equiv\mathcal{N}(T)=\int\mathcal{D}p\; e^{-\frac{\mathcal{E}}{2}\int\limits_0^Td\tau\;p^2(\tau)}\,.
\end{align}

With this path integral definition of the real part of fermion effective action,  one can begin to define currents (and products thereof). One obtains for instance the 
vector current $\langle j_\mu^V(y)\rangle$ to be\footnote{Note that this expression is still written in an Euclidean formulation. The continuation of this and like expressions to real-time is straightforward as we will show in section \ref{sec:ckt}.}
\begin{align}\label{eq:currentreal}
\langle j_\mu^V(y)\rangle&=\frac{\delta\Gamma_\mathbb{R}}{\delta A_\mu(y)}=-\frac{i}{8}\int\limits_0^\infty\frac{dT}{T}\mathcal{N}\int\limits_P\mathcal{D}x\int\limits_{AP}\mathcal{D}\psi
 \;j_{\mu}^{V,cl}\left( e^{-\int\limits_0^Td\tau\;\mathcal{L}
_L(\tau)}+ e^{-\int\limits_0^Td\tau\;\mathcal{L}_R(\tau)}\right),\\
j_{\mu}^{V,cl}&\equiv \int\limits_0^Td\tau\;[\mathcal{E}\psi_\nu\psi_\mu\partial_\nu-\dot{x}_\mu]\delta^4\left(x(\tau)-y\right) \label{eq:realvariation}
\end{align}
It can be easily shown that 
\begin{align}\label{eq:currentcons}
\partial_\mu \langle j_\mu^V\rangle = 0 \qquad\Leftrightarrow\qquad \partial_\mu  j_\mu^{V,cl} = 0.
\end{align}
In proving these relations, we first used the definition of the total derivative for the divergence of the first term of \Eq{eq:realvariation}, employed our knowledge of the boundary terms and used 
that
\begin{align}
\int\limits_0^Td\tau\; \dot{x}_\mu\frac{\partial}{\partial y_\mu}\delta^4(x(\tau)-y)=-\int\limits_0^Td\tau\; \frac{d}{d\tau}\delta^4(x(\tau)-y)=0.
\end{align}
The second term in the four divergence of \Eq{eq:realvariation} vanishes by the anti-symmetry of the Grassmann variables, when interchanging the $y$ and $x(\tau)$ derivatives. We note further that the world-line description provides us with a natural regularization as discussed in \cite{FiorenzoBookNew}, whereby 
$T\rightarrow 0$ represents the ultraviolet limit of the effective action and $T\rightarrow \infty$ is related to the infrared limit. 
%
%
\subsection{Imaginary Part}\label{sec:WLframework:imag}
\subsubsection{World-line representation of the phase of the fermion determinant}
In this section, we will derive a path integral representation of the imaginary part of the fermion effective action, as defined in \Eq{eq:realimagpartseff}. As noted in  \cite{AlvarezGaume:1983ig}, the absolute value of the phase of 
the fermion determinant is not well defined (for fermions in a complex representation). On the other hand, \textit{variations} or \textit{relative} phases (variation writh regards to an external parameter), are unambiguous. In the world-line framework, the fact that the absolute value of the phase in the fermion determinant is ill-defined is 
reflected by the lack of a heat kernel regularization for the imaginary part of the effective action -- the latter is only possible when the action breaks axial symmetry explicitly. 

We proceed with our discussion by expressing the relation between the phase of the fermion determinant and the corresponding imaginary part of the resulting effective action as
\begin{align}
W_\mathbb{I}=-\arg\det[\theta]\,,\label{eq:imag}
\end{align}
where $\theta$ is defined in \Eq{eq:effectiveaction}. Again, extending the dimensionality of $\theta$,  we can write the above as  
\begin{align}
W_\mathbb{I}=-\frac{1}{2}\arg\det[\Omega],\qquad
\Omega=\begin{pmatrix}
0 &\theta\\
\theta & 0 
\end{pmatrix}\,,
\end{align}
where $\Omega$, which is an $8\times 8$ dimensional matrix which reads 
\begin{align}
\Omega=\Gamma_\mu(p_\mu-A_\mu)-i\Gamma_7\Gamma_\mu\Gamma_5\Gamma_6 B_\mu\, .
\end{align}
The Gamma matrices are those defined previously in \Eq{eq:gammamatrices} and \Eq{eq:gammamatrix7}. A lengthy derivation, that includes a further doubling of dimensions -- discussed in 
\cite{D'Hoker:1995ax,D'Hoker:1995bj} in full detail -- results in the expression
\begin{align}\label{eq:imagoperator}
-i W_\mathbb{I}=\frac{1}{4}\text{Tr}\log{\tilde{\Omega}}-\frac{1}{4}\text{Tr}\log{\tilde{\Omega}^\dagger}\,,
\end{align}
where $\tilde{\Omega}$ is given as
\begin{align}\label{eq:omega1}
\tilde{\Omega}=\frac{1}{2}(\tilde{\Sigma}-\tilde{\Sigma}^c)i\Gamma_6\Gamma_7+\frac{i}{2}\Gamma_5\Gamma_6\Gamma_7\chi(\tilde{\Sigma}-\tilde{\Sigma}^c)
i\Gamma_6\Gamma_7\,,
\end{align}
with
\begin{align}
\tilde{\Sigma}=\Gamma_\mu(p_\mu-\mathcal{A}_\mu),\qquad \chi=
\begin{pmatrix}
1 & 0 \\
0 & -1
\end{pmatrix}\,.
\end{align}
We note that $\tilde{\Sigma}^c$ is the chiral conjugate of $\tilde{\Sigma}$, by setting $B\rightarrow -B$. This expression allows one to represent the phase of the fermion determinant as the trace of logarithms, as previously.
The crucial difference to the real part however is that \Eq{eq:omega1} does not permit a path integral representation analogous to \Eq{eq:tr}. This is principally because the operator $\tilde{\Omega}$ does not have a 
positive-definite spectrum, respectively heat kernel expression. 

Nevertheless this obstacle is overcome by a trick due to D'Hoker and Gagn\'e~\cite{D'Hoker:1995ax,D'Hoker:1995bj}. 
Inserting an auxilliary parameter~$\alpha$, \Eq{eq:imagoperator} can be written as
\begin{align}
-i W_\mathbb{I}&=\frac{1}{4}\int\limits_{-1}^1d\alpha\;\frac{\partial}{\partial\alpha}\left(\text{Tr}\log\left[\frac{1}{2}(\tilde{\Omega}+\tilde{\Omega}^\dagger)+\frac{\alpha}{2}(\tilde{\Omega}-\tilde{\Omega}^\dagger) \right]\right)\nonumber\\
&=\frac{1}{4}\int\limits_{-1}^1d\alpha\;\text{Tr}\left\{ \frac{\tilde{\Omega}-\tilde{\Omega}^\dagger}{(\tilde{\Omega}+\tilde{\Omega}^\dagger)
+\alpha(\tilde{\Omega}-\tilde{\Omega}^\dagger)}\right\}\,.
\end{align}
Symmetrizing this expression with respect to $\alpha$ gives
\begin{align}
\frac{1}{4}\int\limits_{-1}^1d\alpha\;\text{Tr}\left\{\frac{\tilde{\Omega}^2
-\tilde{\Omega}^{\dagger 2}}{(\tilde{\Omega}+\tilde{\Omega}^\dagger)^2+2\alpha[\tilde{\Omega},\tilde{\Omega}^\dagger]-\alpha^2(\tilde{\Omega}^2-\tilde{\Omega}^{\dagger 2})}\right\}\,.
\end{align}
There is an identity that ensures that the denominator of this expression is positive-definite and admits a heat kernel regularization~\cite{D'Hoker:1995ax}. Keeping the numerator however separate and defining it as 
\begin{align}\label{eq:traceinsertion}
\hat{M}\equiv \tilde{\Omega}^2-\tilde{\Omega}^{\dagger 2}\,,
\end{align}
in analogy to section \ref{sec:WLframework:real},  the imaginary part of the effective action can be expressed as 
\begin{align}\label{eq:heatkernelimag}
W_\mathbb{I}=\frac{i\mathcal{E}}{64}\int\limits_{-1}^1 d\alpha\int\limits_0^\infty dT\;\text{Tr}\left\{ \hat{M}e^{-\frac{\mathcal{E}}{2}T\tilde{\Sigma}^2_{(\alpha)}}\right\}\,.
\end{align}

Remarkably the matrix $\tilde{\Sigma}^2_{(\alpha)}$ coincides with $\tilde{\Sigma}^2$ that enters \Eq{eq:tr}, albeit with the replacement of the axial-vector field therein by ~$B\rightarrow \alpha B$. This result permits us to properly interpret $\alpha$  as the parameter regulating chiral symmetry breaking in the effective action. The values $\alpha=\pm 1$ correspond to the coupling of gauge fields to left- (right-) handed particles. 
Since \Eq{eq:heatkernelimag} contains an continuous integral over $\alpha$, chiral symmetry is necessarily broken for $\alpha \neq \pm 1$. There is a trace insertion \Eq{eq:traceinsertion}, in \Eq{eq:heatkernelimag} 
that is absent in the real part of the effective action\footnote{This contribution is analogous to the $\gamma_5$ insertion in ``textbook" discussions of the anomaly~\cite{Polyakov:1987ez}.}. This can be split into two contributions,
\begin{align}\label{eq:wlinsertiondef}
\hat{M}=\Gamma_7\Lambda,\qquad\quad
\Lambda= \Lambda^{(1)}+\Lambda^{(2)},
\end{align}
which are given as
\begin{align}\label{eq:insertionlambdas}
\Lambda^{(1)} &\equiv  2\Gamma_5\Gamma_6[\partial_\mu,B_\mu]\,\mathbb{I}_{2},\nonumber\\
\Lambda^{(2)} &\equiv [\Gamma_\mu,\Gamma_\nu]\{\partial_\mu,B_\nu \}\Gamma_5\Gamma_6\, \mathbb{I}_{2}\,.
\end{align}
Both contributions are linear in the axial-vector field $B$ and further are diagonal in the (two-dimensional) field representation space introduced in \Eq{eq:internalfieldrep}. 
Just as in the case for the real part of the effective action, the coherent state basis \Eq{eq:coherent1} can be used to present the trace in \Eq{eq:heatkernelimag} as follows
\begin{align}\label{eq:imag:trace}
\text{Tr}\left\{ \hat{M}e^{-\frac{\mathcal{E}}{2}T\tilde{\Sigma}^2_{(\alpha)}}\right\}=
\int d^4z\; d^3\theta \;\langle z, -\theta |\hat{M}e^{-\frac{\mathcal{E}}{2}T\tilde{\Sigma}^2_{(\alpha)}} | z, \theta\rangle.
\end{align}
Here the (trivial) sub-trace over the two-dimensional field representation space is implicit. From \Eq{eq:imag:trace}, a path integral representation can be found; however the insertion of the operator $\hat{M}$ in the trace requires  care in the discretization of the world-line, more so than for the real case discussed in section \ref{sec:WLframework:real}.

\subsection{The axial-vector current}
Our goal is to derive an expression for the global axial-vector current, defined as 
\begin{align}\label{eq:def:axialcurrent}
\langle j_\mu^5(y) \rangle\equiv \frac{i\delta W_\mathbb{I}[A,B]}{\delta B_\mu(y)}\Big|_{B=0}\,.
\end{align}
We will subsequently derive the famous anomaly equation in our approach demonstrating that this current is not conserved. \Eq{eq:def:axialcurrent}
can be written as 
\begin{align}\label{eq:variation}
\langle j_\mu^5(y) \rangle\equiv \frac{i\delta W_\mathbb{I}}{\delta B_\mu(y)}\Big|_{B=0}=-\frac{\mathcal{E}}{64}\int\limits_0^1d\alpha\int\limits_0^\infty dT\;\text{Tr}
\left\{ \frac{\delta\hat{M}}{\delta B_\mu(y)}e^{-\frac{\mathcal{E}}{2}T\tilde{\Sigma}^2_{(\alpha)}}\right\}_{B=0}
=-\frac{\mathcal{E}}{32}\int\limits_0^\infty dT\;\text{Tr}
\left\{ \frac{\delta\hat{M}}{\delta B_\mu(y)}e^{-\frac{\mathcal{E}}{2}T\tilde{\Sigma}^2}\right\}_{B=0}\,.
\end{align}
Note that the variation of the exponential with respect to $B^\mu$ does not contribute when $B^\mu$ is set to zero. The surviving expression above contains both terms in \Eq{eq:wlinsertiondef}. We will discuss both separately. The trace in \Eq{eq:variation} is written as 
\begin{align}\label{eq:traceinsertion:variation}
\text{Tr}
\left\{ \frac{\delta\hat{M}}{\delta B_\mu(y)}e^{-\frac{\mathcal{E}}{2}T\tilde{\Sigma}^2}\right\}_{B=0}&=\text{tr}\int d^4 x^0\;d^3\theta^0\; \langle x^0,-\theta^0 | \;
\frac{\delta\hat{M}}{\delta B_\mu(y)}e^{-\frac{\mathcal{E}}{2}T\tilde{\Sigma}^2}\; | x^0, \theta^0 \rangle\nonumber\\
&=\text{tr}\int d^4 x^0\;d^3\theta \;\langle x^0,\theta^0 | \;
\frac{\delta\Lambda}{\delta B_\mu(y)}\;e^{-\frac{\mathcal{E}}{2}T\tilde{\Sigma}^2} \;| x^0, \theta^0 \rangle
\end{align}
Here, in going from the first line to the second, we made use of $\langle -\theta|\Gamma_7 = \langle \theta|$. 
In particular, $\Gamma_7$ can be shown to be the equivalent of $(-1)^F$, where $F$ is the fermion number operator defined from the coherent states in \Eq{eq:coherent1}~(c.f. \cite{D'Hoker:1995ax}).

This has important consequences: due to this world-line insertion, the path integral representation of the imaginary part of the fermion effective action will contain an integration over Grassmannian variables with \textit{periodic} boundary conditions. Consequently, fermionic zero modes arise, which would not be present otherwise. By insertion of complete sets of states, \Eq{eq:variation} can be written as
\begin{align}\label{eq:prodmatrix}
\langle j_\mu^5(y) \rangle = -\frac{\mathcal{E}}{32}\int\limits_0^\infty dT\,\text{tr}\int d^4 x^0 d^3\theta^0 d^4 x^N d^3\theta^N \;\langle x^0,\theta^0 | \;
\frac{\delta\Lambda}{\delta B_\mu(y)}| x^N,\theta^N\rangle
\langle x^N,\theta^N|\;e^{-\frac{\mathcal{E}}{2}T\tilde{\Sigma}^2} \;| x^0, \theta^0 \rangle.
\end{align}
Both matrix elements in the r.h.s of this expression can be treated separately. We begin with the matrix element containing the exponential. In analogy with previous derivations, we get
\begin{align}\label{eq:pathint}
\langle x^N,\theta^N|&\;e^{-\frac{\mathcal{E}}{2}T\tilde{\Sigma}^2} \;| x^0, \theta^0 \rangle=
-\int \left(\prod\limits_{k=1}^{N-1}d^4 x_k\right)\left(\prod\limits_{k=1}^{N}\frac{d^4p_k}{(2\pi)^4}\right)
\left(\prod\limits_{k=1}^{N-1}d^3 \theta_k\right)\left(\prod\limits_{k=1}^{N}d^3 \bar{\theta}_k\right)\nonumber\\&
\times\exp \Big\{-\Delta\sum\limits_{k=1}^{N}\big[ 
-ip_\mu^k\frac{(x^k_\mu-x^{k-1}_\mu)}{\Delta}+\frac{\mathcal{E}}{2}\left(p_\mu^k-A_\mu[\bar{x}^k] \right)^2
\quad - \frac{(\theta^k_r-\theta^{k-1}_r)}{\Delta}\bar{\theta}^k_r
+\frac{i\mathcal{E}}{2}\psi^k_\mu
F_{\mu\nu}[\bar{x}^k]\psi_\nu^{k-1}
\big]\Big\}\,.
\end{align}
We now proceed to evaluate the matrix element in \Eq{eq:prodmatrix} that contains the world-line insertion. From \Eq{eq:wlinsertiondef}, the latter can be split into separate parts. 
We begin our discussion with $\Lambda^{(1)}$, which gives 
\begin{align}\label{eq:insert1}
\langle x^0,\theta^0| \frac{\delta\Lambda^{(1)}}{\delta B_\mu(y)}| x^N,\theta^N\rangle=2\left( \frac{\partial}{\partial x^0_\mu}\delta(x^0-y)  \right)\delta(x^0-x^N)\langle \theta^0|\Gamma_5\Gamma_6| \theta^N\rangle\,.
\end{align}
The second world-line insertion $\propto\Lambda^{(2)}$ is similarly
\begin{align}\label{eq:insert2}
\langle x^0,\theta^0| \frac{\delta\Lambda^{(2)}}{\delta B_\mu(y)}| x^N,\theta^N\rangle
=\Big\{ \left(\frac{\partial}{\partial x^0_\nu}\delta(x^0-y)\right)\delta(x^0-x^N)+2\left(\frac{\partial}{\partial x_\nu^0}\delta(x^0-x^N)\right)\delta(\bar{x}^0-y) \Big\}
\langle \theta^0|[\Gamma_\nu,\Gamma_\mu]\Gamma_5\Gamma_6| \theta^N\rangle\,.
\end{align}
Adding together \Eq{eq:insert1} and \Eq{eq:insert2}, multiplying it with the matrix element in \Eq{eq:pathint}, and inserting this expression in the r.h.s of \Eq{eq:prodmatrix}, 
gives us the complete world-line expression for the anomalous axial vector current.

\subsubsection{Derivation of the axial anomaly}
To determine the anomaly equation, we need to compute $\partial_\mu (\delta iW_\mathbb{I}/B_\mu(y))_{B=0}$. We should mention here at the outset that only
\Eq{eq:insert1} contributes to the anomalous non-conservation of the axial-vector current, while \Eq{eq:insert2} does not; this statement is illustrated in appendix \ref{sec:app:anomalydetails}. One thus obtains 
\begin{align}
\partial_\mu \langle j_\mu^5(y) \rangle = \partial_\mu\frac{i\delta W_\mathbb{I}}{\delta B_\mu(y)}\Big|_{B=0}=
-\frac{\mathcal{E}}{32}\int\limits_0^\infty dT\;\partial_\mu\text{Tr}&\left( \Gamma_7 \frac{\delta\Lambda^{(1)}}{\delta B_\mu(y)} e^{-\frac{\mathcal{E}}{2}T\tilde{\Sigma}^2} \right)
\end{align}
where the trace is now written as
\begin{align}\label{eq:combinmag}
&\partial_\mu\text{Tr}\left( \Gamma_7 \frac{\delta\Lambda^{(1)}}{\delta B_\mu(y)} e^{-\frac{\mathcal{E}}{2}T\tilde{\Sigma}^2} \right)=-8\int\left(\prod\limits_{l=0}^{N-1}d^4x^l \right)\left(\prod\limits_{i=1}^N\frac{d^4p^i}{(2\pi)^4}\right)\left(\prod\limits_{j=0}^N d^3\theta^j 
d^3\bar{\theta}^j\right)
\left(\frac{\partial^2}{\partial y_\mu \partial x^0_\mu}\delta(x^0-y) \right)\psi_5^0 \psi^N_6 \nonumber\\
&\times\exp\Big\{ -\Delta\sum\limits_{k=1}^N\Big[ 
-ip_\alpha^k\frac{(x_\alpha^k-x_\alpha^{k-1})}{\Delta}-\frac{(\theta_r^k-\theta_r^{k-1})}{\Delta}+\frac{\mathcal{E}}{2}\left(p_\alpha^k-A_\alpha(\bar{x}^k) \right)^2+\frac{i\mathcal{E}}{2}\psi_\alpha^k\psi_\beta^{k-1}F_{\alpha\beta}(\bar{x}^k)
\Big]+(\theta^0_r-\theta_r^N)\bar{\theta^0_r}\Big\}\,.
\end{align}
We have made use of \Eq{eq:psitheta} to write this expression in a compact form. We can now follow the same procedure as for the real part and complete the squares for the  $p_k$ ($k=1,\dots,N$) integration\footnote{This of course does not effect the integration variables in the representation of the world-line insertion. We emphasize this point, because in the compact notation in \cite{D'Hoker:1995ax}, this procedure is unclear and may cause confusion.} 
\begin{align}
p_\alpha^k\rightarrow p_\alpha^k-A_\alpha[\bar{x}^k]-i\frac{(x^k_\alpha-x^{k-1}_\alpha)}{\Delta\mathcal{E}} \,.
\end{align}
We then find
\begin{align}\label{eq:combinmagSIMPLIFIED}
\partial_\mu\text{Tr}\left( \Gamma_7 \frac{\delta\Lambda^{(1)}}{\delta B_\mu(y)} e^{-\frac{\mathcal{E}}{2}T\tilde{\Sigma}^2} \right)=-8&\left\{\int\prod\limits_{l=1}^N\frac{d^4p^l}{(2\pi)^4} e^{-\Delta\sum\limits_{k=1}^N\frac{\mathcal{E}}{2}(p^k)^2} \right\}
\nonumber\\&\times \int\left(\prod\limits_{i=0}^{N-1}d^4x^i \right)\left(\prod\limits_{j=0}^N d^3\theta^j d^3\bar{\theta}^j\right)
\left(\frac{\partial^2}{\partial y_\mu \partial x^0_\mu}\delta(x^0-y) \right)\psi_5^0 \psi^N_6 
\exp{\left\{ -\Delta\sum\limits_{k=1}^N \mathcal{L}^k\right\}}\,.
\end{align}
The exponential factor in the latter expression is
\begin{align}\label{eq:expfactor}
\exp{\left\{ -\Delta\sum\limits_{k=1}^N \mathcal{L}^k\right\}}\equiv\exp\Big\{ -\Delta\sum\limits_{k=1}^N\Big[ 
\frac{1}{2\mathcal{E}}\frac{(x_\alpha^k-x_\alpha^{k-1})^2}{\Delta^2}-\frac{(\theta_r^k-\theta_r^{k-1})}{\Delta}-i\frac{(x_\alpha^k-x_\alpha^{k-1})}{\Delta}A_\alpha(\bar{x}^k)\nonumber\\+\frac{i\mathcal{E}}{2}\psi_\alpha^k\psi_\beta^{k-1}F_{\alpha\beta}(\bar{x}^k)
\Big]+(\theta^0_r-\theta_r^N)\bar{\theta^0_r}
\Big\}\,.
\end{align}
By means of partial integration $\frac{\partial^2}{\partial y_\mu \partial x_\mu^0}\delta(x^0-y)=-\delta(x^0-y) \frac{\partial^2}{\partial x^0_\mu \partial x_\mu^0}$, we get
\begin{align}\label{eq:derivanomnoncons}
\frac{\partial^2}{\partial x^0_\mu \partial x_\mu^0}\exp{\left\{ -\Delta\sum\limits_{k=1}^N \mathcal{L}^k\right\}}=\Big[ -\frac{8}{\mathcal{E}\Delta}-2i\left(\frac{\partial}{\partial x_\mu^0}A_\mu(\bar{x}^1)
-\frac{\partial}{\partial x_\mu^0}A_\mu(\bar{x}^0) \right) 
-i(x^1_\alpha-x^0_\alpha)\frac{\partial^2}{\partial x_\mu^0\partial x_\mu^0}A_\alpha(\bar{x}^1)\nonumber\\-i(x^0_\alpha-x^N_\alpha)\frac{\partial^2}{\partial x_\mu^0\partial x_\mu^0}A_\alpha(\bar{x}^0)+O(\Delta)
\Big]e^{ -\Delta\sum\limits_{k=1}^N \mathcal{L}^k}\longrightarrow -\frac{8}{\mathcal{E}\Delta} \exp{\left\{ -\Delta\sum\limits_{k=1}^N \mathcal{L}^k\right\}}\,,
\end{align}
where the leading terms in the limit of $k\rightarrow \tau$, $\Delta\rightarrow 0$ are kept. In the continuum limit and setting $B=0$ we have
\begin{align}\label{eq:derivBnull}
\partial_\mu\frac{i\delta W_\mathbb{I}}{\delta B_\mu(y)}&=2\int\limits_0^\infty dT \mathcal{N}(T)\;\int\limits_{P} Dx \int\limits_{P} D\psi\; (\psi_5\psi_6)(0)\delta(x(0)-y)\nonumber\\
&\times\exp\left\{ -\int\limits_0^T d\tau\;\frac{1}{2\mathcal{E}}\dot{x}^2-i\dot{x}_\alpha A_\alpha(x)-\frac{1}{2}\psi_a\dot{\psi}_a+\frac{i\mathcal{E}}{2}\psi_\mu F_{\mu\nu}\psi_\nu \right\}\,,
\end{align}
where $(\psi_5\psi_6)(0)$ is an insertion of the respective Grassman variables at world-line ``time" $\tau=0$.

We  will now find an analytic solution for \Eq{eq:derivBnull}. To this end, we remark that, as illustrated above,
both anti-commuting as well as commuting world-line variables are defined with periodic boundary conditions. We can therefore write both respectively  as a sum of a zero mode and a proper time dependent contribution 
\begin{align}\label{eq:zeromodesdef1}
x_\mu(\tau)&=\bar{x}_\mu+x'_\mu(\tau)\,,\\
\psi_a(\tau)&=\bar{\psi}_a+\psi'_a(\tau)\,,
\end{align}
where the zero modes are defined to be 
\begin{align}
\bar{x}_\mu\equiv\int\limits_0^T d\tau \;x_\mu(\tau)\qquad\qquad \bar{x}_\mu={x}_\mu(0)={x}_\mu(T)\\
\bar{\psi}_a\equiv\int\limits_0^T d\tau \; \psi_a(\tau)\qquad\qquad\bar{\psi}_a={\psi}_a(0)={\psi}_a(T)
\end{align}
and similarly for $\psi_5,\psi_6$. The latter two fields can be trivially integrated in \Eq{eq:derivBnull}. The result is
\begin{align}
\int\limits_{P} d\psi\;\psi \,e^{-\frac{1}{2}\int\limits_0^Td\tau\;\psi\dot{\psi}}=
\int d\psi^0d\psi'\;(\psi^0+\psi ')\,e^{-\frac{1}{2}\int\limits_0^Td\tau\;\psi'\dot{\psi'}}=1\,,
\end{align}
where $\psi$ stands for either $\psi_5,\psi_6$. We will henceforth define the remaining integral measure as 
$\mathcal{D}\psi\equiv \prod_{\mu=1}^4\mathcal{D}\psi_\mu$. The result can be compactly summarized as
\begin{align}\label{eq:contpathimag}
\partial_\mu\frac{i\delta W_\mathbb{I}}{\delta B_\mu(y)}\Big|_{B=0}=2\int\limits_0^\infty dT&\; \mathcal{N}(T)\;\int D\bar{x}Dx' \int D\bar{\psi} D\psi'\;\delta(\bar{x}-y)\nonumber\\&\times
\exp\left\{ -\int\limits_0^T d\tau\;\frac{1}{2\mathcal{E}}\dot{x'}^2-i\dot{x'}_\alpha A_\alpha(x)-\frac{1}{2}\psi'_\mu\dot{\psi}'_\mu+\frac{i\mathcal{E}}{2}\psi'_\mu F_{\mu\nu}\psi'_\nu
+\frac{i\mathcal{E}}{2}\bar{\psi}_\mu F_{\mu\nu}\bar{\psi}_\nu\right\}\,,
\end{align}
where the normalization $\mathcal{N}$ is as in \Eq{eq:normalization}.

Because this normalization has a strong power law dependence on $1/T$~\cite{Polyakov:1987ez}, the path integral receives its largest contributions from
$T\rightarrow 0$. As the non-zero modes in \Eq{eq:zeromodesdef1} can be expanded in terms of eigenmodes with frequencies $T^{-1}$,  higher modes do not contribute to the $T\rightarrow 0$
limit. It is therefore sufficient to expand the integrand around the zero modes, keeping non-zero modes only up to quadratic order. To evaluate this, it is 
convenient to use Fock-Schwinger gauge\footnote{This procedure was discussed in detail in \cite{AlvarezGaume:1983ig,AlvarezGaume:1983at}.}, centered around $\bar{x}$,
which is defined by
\begin{align}\label{eq:FockSchwingerGauge}
x'_\mu(\tau)A_\mu(\bar{x}+x'(\tau))=0.
\end{align}
This expression can be formally solved for $A$, which results in
\begin{align}
A_\mu(\bar{x}+x')=x'_\nu\int\limits_0^1d\eta\;\eta \;F_{\nu\mu}(\bar{x}+\eta x')=x'_\nu \int\limits_0^1d\eta\;\eta\;\exp\left({\eta\, x'_\alpha \partial_\alpha}\right)F_{\nu\mu}(\bar{x})
\end{align}
As we are expanding around the zero modes, it is sufficient to expand
\begin{align}
A_\mu(\bar{x}+x')=\frac{1}{2}{x'}_\nu F_{\nu\mu}(\bar{x})+\frac{1}{3}{x'}_\nu {x'}_\rho\partial_\nu F_{\rho\nu}(\bar{x})+\dots.
\end{align}
In fact, we only need to keep
\begin{align}
A_\mu(x)\approx\frac{1}{2}F_{\mu\nu}(\bar{x})x'_\nu\,.
\end{align}
Exploiting Fock-Schwinger gauge thusly, \Eq{eq:contpathimag} can be brought into the appealing form\footnote{As will become clear from our derivation below, \Eq{eq:contpathimag2} carries in fact an $N=1$ supersymmetry,  turning bosonic into fermionic variables and vice-versa. Details are
given in appendix~\ref{app:susy}.},
\begin{align}\label{eq:contpathimag2}
\partial_\mu\frac{i\delta W_\mathbb{I}}{\delta B_\mu(y)}\Big|_{B=0}=2\int\limits_0^\infty dT& \mathcal{N}(T)\;\int D\bar{x}Dx' \int D\bar{\psi} D\psi'\;\delta(\bar{x}-y)\nonumber\\&\times
\exp\left\{ -\int\limits_0^T d\tau\;\frac{1}{2\mathcal{E}}\dot{x'}^2-\frac{i}{2}x'_\mu F_{\mu\nu}\dot{x}'_\nu-\frac{1}{2}\psi'_\mu\dot{\psi}'_\mu+\frac{i\mathcal{E}}{2}\psi'_\mu F_{\mu\nu}(\bar{x})\psi'_\nu 
+\frac{i\mathcal{E}}{2}\bar{\psi}_\mu F_{\mu\nu}(\bar{x})\bar{\psi}_\nu\right\}\,.
\end{align}
 We proceed by performing the (quadratic) non-zero mode integration in \Eq{eq:contpathimag2}. The results of performing these integrals are \cite{Schubert:2001he}
\begin{align}\label{eq:det1}
\int\mathcal{D}x'\;\exp{\big\{-\int\limits_0^\infty\big(\frac{{\dot{x'}}^2}{4} - \frac{i}{2}x'_\mu F_{\mu\nu}\dot{x}'_\nu  \big) \big\}}&=\text{Det}^{\prime\,-\frac{1}{2}}
\left( -\frac{d^2}{d \tau^2}+2i F\frac{d}{d\tau} \right)\nonumber\\
&=\frac{1}{(4\pi T)^2}\text{Det}^{\prime\,-\frac{1}{2}}\left( 1-2iF\left(\frac{d}{d\tau} \right)^{-1} \right)=\frac{1}{(4\pi T)^2}\text{det}^{-\frac{1}{2}}\left(\frac{\sin(FT)}{FT} \right)
\end{align}
and
\begin{align}\label{eq:det2}
\int \mathcal{D}\psi'\;\exp{\big\{ -\int\limits_0^T\big(  \frac{1}{2}\psi'_\mu\dot{\psi'}_\mu+i\psi'_\mu F_{\mu\nu}\psi'_\nu        \big)    \big\}}=\text{det}^{\frac{1}{2}}\left(\frac{\sin(FT)}{FT} \right)\,.
\end{align}
Here $\text{Det}^\prime$ indicates the determinant acting on the space of variables \textit{sans} the zero modes, while $\text{det}$ is defined on the reduced space on which the gauge field tensor $F$ is defined.
Due to the N=1 supersymmetry of \Eq{eq:contpathimag2}, the fermionic and bosonic integrals \Eq{eq:det1} and \Eq{eq:det2} cancel, 
\begin{align}
\int\mathcal{D}x'\mathcal{D}\psi'\;\exp\Big\{-\int\limits_0^\tau d\tau\;
\frac{\dot{x}'^2}{4}+\frac{1}{2}\psi_\mu'\dot{\psi}_\mu'
 -\frac{i}{2}x'_\mu F_{\mu\nu}(\bar{x})\dot{x}'_\nu+i\psi'_\mu F_{\mu\nu}(\bar{x}) \psi'_\nu
\Big\}=\frac{1}{4\pi^2}\frac{1}{4T^2}\,.
\end{align} 
leaving us with the zero mode integration alone:
\begin{align}
\int D\bar{\psi} \; \exp\Big\{-\int\limits_0^\infty d\tau\;i\bar{\psi}_\mu F_{\mu\nu}(\bar{x})\bar{\psi}_\nu\Big\}=\int d^4\bar{\psi} \; 
\exp\Big\{-iT\bar{\psi}_\mu F_{\mu\nu}(\bar{x})\bar{\psi}_\nu\Big\}=-\frac{T^2}{2}\epsilon^{\mu\nu\rho\sigma}F_{\mu\nu}F_{\rho\sigma} \, .
\end{align}
We therefore obtain 
\begin{align}
\partial_\mu\frac{i\delta W_\mathbb{I}}{\delta B_\mu(y)}\Big|_{B=0}&=-\frac{1}{16\pi^2}\int\limits_0^\infty dT \mathcal{N}(T)\;\int D\bar{x}\;\delta(\bar{x}-y)\epsilon^{\mu\nu\rho\sigma}F_{\mu\nu}(\bar{x})F_{\rho\sigma}(\bar{x})
\nonumber\\&=-\frac{1}{16\pi^2}\left( \int\limits_0^\infty dT \mathcal{N}(T)\right)\;\epsilon^{\mu\nu\rho\sigma}F_{\mu\nu}(y)F_{\rho\sigma}(y)\,.
\end{align}
The normalization can be set to unity giving us the well known result
\begin{align}\label{eq:finalanomaly}
\partial_\mu \langle j^5_\mu(y)\rangle\equiv
\partial_\mu\frac{i\delta W_\mathbb{I}}{\delta B_\mu(y)}\Big|_{B=0}=-\frac{1}{16\pi^2}\epsilon^{\mu\nu\rho\sigma}F_{\mu\nu}(y)F_{\rho\sigma}(y).
\end{align}
This is the central result of this section\footnote{We note that commonly in the literature a distinction is being made between covariant and consistent anomalies \cite{Bardeen:1984pm,Bardeen:1969md}. 
In our situation both definitions agree, as is argued in appendix \ref{app:anomalies}. However this distinction is of crucial importance, when deriving non-singlet anomalies or anomalies with both physical vector- 
and axial-vector-background fields present.}. It nicely illustrates that the axial anomaly can be understood as arising from the phase of the fermionic determinant. Unlike many derivations in the literature, we employed a 
a variational technique for the imaginary part of the effective action in an Euclidean formalism. This also confirms that our result for the axial-vector current in \Eq{eq:variation} is robust. The analytic continuation
to Minkowskian metric will be straightforward, albeit the imaginary part of the effective action will have a different interpretation.

In the upcoming section \ref{sec:KineticTheory} we will continue our world-line path integral formulation to real-time  and we will make contact with
the results of \cite{Son:2012wh,Stephanov:2012ki,Son:2012zy,Chen:2013iga,Chen:2014cla} containing a Berry connection. Our very general approach allows one to study the origin and role of any 
geometric phases which arise under certain approximations, such as those corresponding to adiabatic variations in interactions with external fields. We then give an outlook on how a chiral kinetic theory should be constructed, which is equivalent to a saddle point approximation of 
our world-line path integral. In this context, we argue that \Eq{eq:variation} in the pseudo-classical limit provides a consistent definition of the axial vector current and can be used in the construction 
of chiral kinetic extensions of  B\"{o}decker's effective theory~\cite{Bodeker:1998hm,Bodeker:1999ey}.

For completeness, we note that the corresponding definition of the axial-vector current in the continuum formulation of the world-line path integral is given as
\begin{align}
\langle j_5^\mu (y)\rangle = \frac{1}{4}\int\limits_0^\infty dT\,\mathcal{N}\int\limits_P \mathcal{D}x \mathcal{D}\psi\,\delta^{(4)}(x(0)-y)\big\{ [\dot{x}^\mu + \dot{x}^\nu\psi_\mu\psi 
_\nu]\psi_5 \psi_6 \big\}\Big|_{\tau=0}\,\exp{\left(-\int\limits_0^\infty d\tau\, \mathcal{L} \right)},
\end{align}
where with $B=0$ the Lagrangian $\mathcal{L}=\mathcal{L}_L=\mathcal{L}_R$ is given in \Eq{eq:lagrangianexplicit}.
%
%
%
%

\section{Chiral Kinetic Theory}\label{sec:KineticTheory} \label{sec:ckt}
\subsection{Pseudo-classical description of spinning particles}\label{sec:pseudoclassicallimit}
The world-line framework provides a consistent Lorentz covariant description of quantum field theory using the language of first quantization. It is therefore well  suited for a pseudo-classical kinetic description of quantum many-body systems. We will begin our discussion here with the world-line Lagrangian \Eq{eq:realpartLag} continued to Minkowskian metric ($g=\text{diag}[-,+,+,+]$). Henceforth we will consider the coupling of fermions to vector gauge fields and set the auxilliary field $B=0$. We have
\begin{align}\label{eq:lagrangian:RealTime}
\mathcal{L}=\frac{\dot{x}^2}{2\mathcal{E}}+\frac{i}{2}\psi^\mu \dot{\psi}_\mu
+ \frac{i}{2} \psi_5\dot{\psi}_5+ \frac{i}{2} \psi_6\dot{\psi}_6+\dot{x}_\mu A^\mu(x) - \frac{i\mathcal{E}}{2}\psi^\mu F_{\mu\nu}\psi^\nu,
\end{align}
and the corresponding world-line effective action, obtained by the continuation of $W_\mathbb{R}$ from \Eq{eq:realpartpath}, is given by
\begin{align}\label{eq:MINKpartpath}
W=\int\limits_0^\infty\frac{dT}{T}\int\limits_P \mathcal{D}x\int\limits_{AP}\mathcal{D}\psi\;\exp{\Big\{ i\int\limits_0^T d\tau\;\mathcal{L} \Big\}}.
\end{align}
The discussion in section \ref{sec:WLframework:imag} translates into the Minkowskian formulation directly.
The emergence of the anomaly is understood in the Minkowskian formulation as arising from the fact that the path integral measure over the Grassmanian variables in \Eq{eq:MINKpartpath} does not contain zero modes.

The path integral is acompanied by an integration over a world-line of length $T$, which is directly related to the reparametrization invariance of the world-line parameter $\tau\rightarrow\tau'=f(\tau)$. 
In fact, \Eq{eq:MINKpartpath} closely resembles Schwinger's proper time method, albeit in this case the world-line manifold is now an interval in proper time rather than a closed loop. Consequently the world-line lenght $T$ and  the einbein $\mathcal{E}$, 
which is the square root of the determinant of the world-line metric, can also be understood to emerge from a BRST construction (see~\cite{FiorenzoBookNew}). While reparametrization invariance is a gauge symmetry (a redundancy in our
description), it is not related to any symmetry group in the usual sense.  

For particles with spin, yet another physicality condition arises, which is not immediately obvious from \Eq{eq:MINKpartpath}: longitudinal spin components should not be dynamical. This restricts the integral measure
$\mathcal{D}\psi$ to a specific physical hypersurface. In practice, this helicity constraint can be implemented by means of introducing a Lagrange multiplier $\chi$ in the Lagrangian,
\begin{align}
\mathcal{L}\rightarrow\mathcal{L}-i\frac{\dot{x}_\mu\psi^\mu}{2\mathcal{E}}\chi\,.
\end{align}
To illustrate its role,  we  will proceed to the Hamiltonian formulation by defining the conjugate momenta (from \Eq{eq:lagrangian:RealTime})
\begin{align}
\label{eq:conjugate}
p^\mu\equiv\frac{\partial\mathcal{L}}{\partial \dot{x}_\mu}=\pi^\mu+A^\mu\,,\qquad\qquad {\rm with}\qquad\qquad\pi^\mu\equiv\frac{\dot{x}^\mu}{\mathcal{E}}-i\frac{\psi^\mu}{2\mathcal{E}}\chi \, .
\end{align}
The corresponding world line action, equivalent to \Eq{eq:lagrangian:RealTime}, can be written as
\begin{align}\label{eq:hamiltonian1}
S=\int\limits_{0}^{T}d\tau\;\big\{  p_\mu\dot{x}^\mu+\frac{i}{2}[\psi_\mu\dot{\psi}^\mu+\psi_5\dot{\psi}_5+\psi_6\dot{\psi}_6]
-\frac{\mathcal{E}}{2}\pi^2+\frac{i}{2}(\pi_\mu\psi^\mu)\chi-\frac{i\mathcal{E}}{2}\psi^\mu F_{\mu\nu}\psi^\nu \big\}.
\end{align}

The role of $\mathcal{E}$ as a Lagrange multiplier is transparent in the above expression. The constraints that are encoded in \Eq{eq:hamiltonian1} can be easily understood from their quantized counterparts. Promoting the Grassmanian variables to operators in a Hilbert space,
\begin{align}\label{eq:spinquantizationlornez}
\psi_\mu\rightarrow\sqrt{\frac{\hbar}{2}}\gamma_5\gamma_\mu,\qquad \psi_5\rightarrow \sqrt{\frac{\hbar}{2}}\gamma_5\,,
\end{align}
the mass shell condition and the helicity constraint in \Eq{eq:hamiltonian1} correspond to the Klein-Gordon and Dirac operator equations respectively, 
defining the physical subspace $|\Phi\rangle$ of the theory,
\begin{align}\label{eq:physicalsubspace}
\pi^2+i\psi^\mu F_{\mu\nu}\psi^\nu=0\qquad &\Leftrightarrow& \left(\hat{\pi}^2 + i\sigma^{\mu\nu}F_{\mu\nu}\right)|\Phi\rangle =0\qquad&\text{(mass-shell constraint)},\nonumber\\\pi_\mu\psi^\mu=0\qquad&\Leftrightarrow& \gamma_5\gamma_\mu\hat{\pi}^\mu|\Phi\rangle=0\qquad&\text{(helicity constraint)}.
\end{align}
The generalization of \Eq{eq:physicalsubspace} to the massive case is straightforward, as one simply replaces
\begin{align}\label{eq:generalizedconstraint1}
\pi^2+i\psi^\mu F_{\mu\nu}\psi^\nu=0\qquad&\rightarrow\qquad \pi^2+i\psi^\mu F_{\mu\nu}\psi^\nu+m^2=0 \quad&\Leftrightarrow\quad &(\hat{\pi}^2+ i\sigma^{\mu\nu}F_{\mu\nu}+m^2)|\Phi\rangle =0\,, \\\label{eq:generalizedconstraint2}
\pi_\mu\psi^\mu=0\qquad&\rightarrow\qquad\pi_\mu\psi^\mu+m\psi_5=0\quad&\Leftrightarrow\quad &\gamma_5(\gamma_\mu\hat{\pi}^\mu+m)|\Phi\rangle=0\,,
\end{align}
as these then reproduce the \textit{massive} Klein-Gordon and Dirac equations. 

\Eq{eq:generalizedconstraint1} and \Eq{eq:generalizedconstraint2} are not independent. On the operator level, \Eq{eq:generalizedconstraint1}
is the (operator-) squared of \Eq{eq:generalizedconstraint2}, whereas, on the level of the world-line phase space variables $p_\mu,x_\mu,\psi_\mu,\psi_5$, the constraints are part of 
an $N=1$ SUSY algebra, with the supercharge given by \Eq{eq:generalizedconstraint2}. This is discussed further in appendix \ref{app:susy}. In the latter case, both constraints are related by the algebra of Poisson brackets. The action for a spinning massive particle, including both mass-shell and helicity constraints, is then given by 
\begin{align}\label{eq:hamiltonian}
S&=
\int\limits_{0}^{T}d\tau\;\Big\{ p_\mu \dot{x}^\mu+\frac{i}{2}\left[ \psi_\mu\dot{\psi}^\mu+\psi_5\dot{\psi}_5+\psi_6\dot{\psi}_6 \right] -\frac{\mathcal{E}}{2}(\pi^2+m^2)
-\frac{i}{2}\left(\pi_\mu\psi^\mu+m\psi_5\right)\chi- \frac{i\mathcal{E}}{2}\psi^\mu F_{\mu\nu}\psi^\nu \Big\}
\nonumber\\
&\equiv \int\limits_{0}^{T}d\tau\;\Big\{ p_\mu \dot{x}^\mu+\frac{i}{2}\left[ \psi_\mu\dot{\psi}^\mu+\psi_5\dot{\psi}_5+\psi_6\dot{\psi}_6 \right] -H\Big\},
\end{align}
where the Hamiltonian, being merely a sum of constraints, is 
\begin{align}\label{eq:hamilt11}
H=\frac{\mathcal{E}}{2}(\pi^2+m^2+i\psi^\mu F_{\mu\nu}\psi^\nu)+\frac{i}{2}\left(\pi_\mu\psi^\mu+m\psi_5\right)\chi \, .
\end{align}
Since $H$ does not depend on $\psi_6$, the dynamics of the latter is trivial, $\psi_6=const$ and we will drop it from our discussion henceforth. 
\Eq{eq:hamiltonian} serves as our starting point for the determination of the Hamiltonian dynamics of the world-line theory and ultimately leads to the equations of motion in the pseudo-classical (kinetic) limit of the theory. 

The classical limit is not immediately apparent in \Eq{eq:MINKpartpath}  as the $T$ integration obscures its usual interpretation as the saddle point of a path integral with the variables $x,\psi$. However, as described above, the $T$-integration is related to the gauge freedom of the einbein parameter
$\mathcal{E}$. We will  illustrate how this can be dealt with in practice and refer the reader to appendix \ref{sec:saddlepointdef} for further detailed discussion. 

One approach is to perform the $T$ integral in \Eq{eq:MINKpartpath} explictly. In this case, the world-line path integral can be shown to be independent of the value of  the einbein parameter $\mathcal{E}$ and the latter can thus can be fixed to any value. The result of the $T$-integration is a modified single particle action, different from \Eq{eq:lagrangian:RealTime}. The resulting pseudo-classical dynamics can be derived from this modified action, which now permits~\cite{FiorenzoBookNew} only physical degrees of freedom (those satisfying constraint relations) to evolve via the equations of motion.
An alternative approach is as follows: instead of performing the $T$-integral, \Eq{eq:lagrangian:RealTime} might be taken as defining the
the single-particle action directly, albeit explicitly keeping the $T$ integral in \Eq{eq:MINKpartpath}. In this case, $\mathcal{E}$ cannot be fixed and must be treated as a variational parameter.

We will here illustrate both approaches, starting with the first. Fixing $\mathcal{E}=2$ and defining the dimensionless proper time as $u\equiv\tau/T$, \Eq{eq:MINKpartpath} can be written as
\begin{align}
W=\int_0^\infty \frac{dT'}{T'}e^{-iT'} \int \mathcal{D}x \mathcal{D}\psi\,\exp{\Big\{  \frac{i\bar{m}^2}{T'}\int\limits_0^1 du\,\frac{\dot{x}^2}{4} + i\int\limits_0^1du\,
\big[ \frac{i}{2}\left( \psi_\mu\dot{\psi}^\mu + \psi_5\dot{\psi}_5\right) + \dot{x}_\mu A^\mu - \frac{i}{2}\left( \frac{\dot{x}_\mu \psi^\mu}{2} +m\psi_5 \right)\chi \big] \Big\}}\,,
\end{align}
where we further defined $\bar{m}^2\equiv m^2 + i\int_0^1du\, \psi^\mu F_{\mu\nu}\psi^\nu$. Provided the kinetic term is large compared to the interactions, the $T'$ integral can be performed by the stationary 
phase method around the stationary point $T_0'=\bar{m}\sqrt{-\int_0^1du\,\dot{x}^2}$. The result is
\begin{align}
W=\int\mathcal{D}x\mathcal{D}\psi\,\mathcal{\tilde{N}}\,\exp{iS}\,
\end{align}
where $\mathcal{\tilde{N}}\equiv \sqrt{i\pi/2\bar{m}}\left( -\int_0^1 du\,\dot{x}^2 \right)^\frac{1}{4}$, and the corresponding world-line action is
\begin{align}
S=-\bar{m}\sqrt{-\int_0^1 du\,\dot{x}^2}+i\int\limits_0^1du\,\left(\frac{i}{2}\left[ \psi_\mu\dot{\psi}^\mu + \psi_5\dot{\psi}_5+ \psi_6\dot{\psi}_6\right] 
+ \dot{x}_\mu A^\mu - \frac{i}{2}\left[ \frac{\dot{x}_\mu \psi^\mu}{2} +m\psi_5 \right]\chi \right)\,.
\end{align}
Using the abbreviation $Y\equiv\sqrt{-\int_0^1 du\,\dot{x}^2}$, the equations of motion are obtained by varying this (non-local) action,
\begin{align}\label{eq:eomx1}
-\frac{\bar{m}\ddot{x}^\mu}{Y}+\frac{iY}{2\bar{m}}\psi^\alpha\partial^\mu F_{\alpha\beta}\psi^\beta + F^{\mu\nu}\dot{x}_\nu + \frac{i}{4}\dot{\psi}^\mu\chi&=0\,,\\
\dot{\psi}^\mu-\frac{Y}{\bar{m}}F_{\mu\nu}\psi^\nu-\frac{\dot{x}_\mu}{4}\chi&=0\,,\label{eq:eompsimu}\\
\dot{\psi}_5-\frac{m\chi}{2}&=0\,,\label{eq:eompsi5}
\end{align}
while, as noted previously, the dynamics of $\psi_6$ is trivial. These equations of motion, for appropriate choice of $\chi$ (as we shall shortly discuss), provide the covariant generalization of the well known Bargmann-Michel-Telegdi equations~\cite{Bargmann:1959gz}  for spinning particles in external gauge fields. The extension of these equations of motion to include colored degrees of freedom, generalizing thereby the Wong equations~\cite{Wong:1970fu}, was already discussed a long time ago in \cite{Barducci:1976xq}.

As we show in appendix \ref{sec:saddlepointdef}, identical dynamics is obtained in the other approach when $\mathcal{E}$ is treated as a variational parameter and thereby eliminated from the action. This approach will be particularly beneficial when we take the non-relativistic limit of the action. In this variational approach, 
the Euler-Lagrange equations applied to $\mathcal{E}$, using \Eq{eq:lagrangian:RealTime},  give the consistency relation 
\begin{align}\label{eq:consistenycond}
\mathcal{E}=m_R^{-1}\left( z-i\frac{\dot{x}_\mu \psi^\mu}{2\,z}\chi \right),
\end{align}
where $z\equiv\sqrt{-\dot{x}^2}$ and 
\begin{align}
m_R^2=m^2+i\psi^\mu F_{\mu\nu} \psi^\nu.
\end{align}
This consistency relation allows us to eliminate $\mathcal{E}$ by inserting the relation into \Eq{eq:lagrangian:RealTime}. The resulting equation of motions agree with the dynamics in Eqs.~(\ref{eq:eomx1})-(\ref{eq:eompsi5}),
provided the constraints are fulfilled. 

Therefore a saddle point expansion of  \Eq{eq:MINKpartpath} -- under the proviso that all constraints are respected -- provides the correct pseudo-classical limit with the corresponding action given as 
\begin{align}\label{eq:actionT}
S&=\int\limits_{0}^{T}d\tau\;\mathcal{L},
\end{align}
where the Lagrangian in \Eq{eq:lagrangian:RealTime} can now be expressed as 
\begin{align}\label{eq:actionz}
\mathcal{L}\equiv &-\frac{m_R z}{2}\left( 1+\frac{m^2}{m_R^2}\right)+\frac{i}{2}\left(\psi_\mu \dot{\psi}^\mu + \psi_5\dot{\psi}_5 \right) 
- \frac{im_R}{2}\left(\frac{\dot{x}_\mu\psi^\mu}{z}\left[ 1-\frac{m^2}{2m_R^2}\right]+\frac{m}{m_R}\psi_5 \right)\chi
\nonumber\\&+\dot{x}_\mu A^\mu(x) - \frac{i}{2m_R}z\psi^\mu F_{\mu\nu}\psi^\nu.
\end{align}
This Lagrangian, explictly implementing the mass-shell constraint, will serve as the starting point for the discussion of the non-relativistic limit in section \ref{sec:NRadiabatic}.
We can now use \Eq{eq:actionz} to define the  conjugate four-momenta of the constrained phase space; these are 
\begin{align}\label{eq:conjugatemomentaconstraint}
p^\mu\equiv\frac{\partial\mathcal{L}}{\partial\dot{x}_\mu}\,,\qquad\qquad {\rm where}\qquad\qquad\pi^\mu\equiv p^\mu-A^\mu =  m_R u^\mu -  \frac{im_R}{2z}\left(1-\frac{m^2}{2m_R^2} \right)\left[ \psi^\mu + u_\nu\psi^\nu u^\mu \right]\chi,
\end{align}
with the four-velocity is defined as
\begin{align}
u^\mu\equiv \frac{\dot{x}^\mu}{z}\,.
\end{align}
\Eq{eq:conjugatemomentaconstraint} is easily inverted and gives
\begin{align}
\dot{x}^\mu = \frac{z}{m_R}\pi^\mu + \frac{i}{2}\left(1-\frac{m^2}{2m_R^2} \right)\left[  \psi^\mu + \frac{\pi_\nu\psi^\nu \pi^\mu}{m_R^2} \right]\chi
\end{align}
The equations of motion in this setup, respecting all constraints, are completely equivalent to Eqs.~(\ref{eq:eomx1})-(\ref{eq:eompsi5}). This point is illustrated with a specific example in appendix \ref{sec:saddlepointdef}.
We note a few additional points: the Lagrange multiplier $\chi$ is an anti-commuting Lorentz scalar, which means that the structure of expressions that can 
be assigned to it are very restricted~\cite{Berezin:1976eg}. A 
vanishing $\chi=0$ is trivially consistent with this requirement; it turns out the only further choice\footnote{Note that $\chi$ cannot be linear in $\psi$, as this cannot be combined to form a Lorentz invariant. Even powers in $\psi$ result in $\chi$ being a \textit{commuting} variable, rather than an anti-commuting one.} is $\chi\propto \epsilon^{\mu\nu\lambda\sigma}\pi^\mu \psi^\nu\psi^\lambda\psi^\sigma$~\cite{Berezin:1976eg}.

Our considerations here are essential ingredients in deriving a consistent relativistic chiral kinetic theory. The explicit derivation of this kinetic framework is fairly involved and will be left to forthcoming work~\cite{follow-up}.  In the following subsection, we will discuss the role of $\chi$ in more detail and we shall fix it explicitly. 
Our focus in \ref{sec:NRadiabatic} will however be on the non-relativistic reduction of \Eq{eq:actionz}--with the helicity constraint imposed. We will comment on some interesting features of the corresponding  kinetic theory that are complementary to those discussed in our recent letter \cite{Mueller:2017lzw}. 
\subsection{The non-relativistic limit}\label{sec:NRadiabatic}
In this subsection,  we shall derive the non-relativistic limit of the single particle action defined by \Eq{eq:actionz}. We will carefully discuss the role of the mass-shell and helicity constraints and their related
Lagrange-multipliers. Based on an adiabatic approximation of our result, we make contact with the geometric action put forward by \cite{Son:2012wh,Stephanov:2012ki}. As in those works, we showed in our accompanying letter \cite{Mueller:2017lzw} how a Berry term arises in the massive  non-relativistic and adiabatic limit. However in \cite{Mueller:2017lzw}, we only considered a massive system with a small or vanishing chemical potential. We will extend the discussion here to a system with a large chemical potential. 

We will begin by writing \Eq{eq:actionz} with all factors of $c$ specified: 
\begin{align}\label{eq:actionz1}
\mathcal{L}= &-\frac{m_R c\,z}{2}\left( 1+\frac{m^2}{m_R^2}\right)+\frac{i}{2}\left(\psi_\mu \dot{\psi}^\mu + \psi_5\dot{\psi}_5 \right) 
- \frac{im_R\,c}{2}\left(\frac{\dot{x}_\mu\psi^\mu}{z}\left[ 1-\frac{m^2}{2m_R^2}\right]+\frac{m}{m_R}\psi_5 \right)\chi
\nonumber\\&+\frac{\dot{x}_\mu A^\mu(x)}{c} - \frac{i}{2m_Rc}z\psi^\mu F_{\mu\nu}\psi^\nu.
\end{align}
The non-relativistic limit can be derived systematically in an expansion of the particle's velocity over the speed of light. The adiabatic limit corresponds to taking the interaction energy of the particle with the external electromagnetic fields to be small relative to its rest energy. To proceed further in deriving these limits from the relativistic Lagrangian, we choose, without loss of generality, $\chi=0$. It follows thence from \Eq{eq:eompsi5} that $\dot{\psi_5}=0$ and hence $\psi_5=const$. 

We will next use the supersymmetric properties of the world-line action (discussed in appendix \ref{app:susy})
\begin{align}
\psi_\mu\rightarrow\psi_\mu+\frac{\dot{x}_\mu}{\sqrt{-\dot{x}^2}}\,\eta\,;\qquad
\psi_5\rightarrow\psi_5+\eta\,;\qquad
x_\mu\rightarrow x_\mu+i\frac{\psi_\mu\eta}{m}\,,
\end{align}
where $\eta$ is an anticommuting parameter generating a $N=1$ supersymmetric transformation. Since $\psi_5=const$, we can perform a time-independent transformation such that $\psi_5=0$. Thereby eliminating $\psi_5$ from the dynamics entirely,  the Lagrangian can be written as 
\begin{align}\label{eq:lagnrangianstartingpoint}
\mathcal{L}= -\frac{m_R c\,z}{2}\left( 1+\frac{m^2}{m_R^2}\right)+\frac{i}{2}\left(\boldsymbol{\psi} \dot{\boldsymbol{\psi}}- \psi_0\dot{\psi}_0 \right) 
+\frac{\dot{x}_\mu A^\mu(x)}{c} - \frac{i}{m_Rc}z\,\psi^0 F_{0i}\psi^i- \frac{i}{2m_Rc}z\,\psi^i F_{ij}\psi^j\,,
\end{align}
This expression does not contain any approximations yet. 

To take the non-relativistic limit, we identify the world line proper  time $\tau$ of  a ``particle", with the physical time $t$ as
\begin{align}
\tau=\frac{ct}{\gamma}=ct\sqrt{1-(d\boldsymbol{x}/dt)^2},\qquad x^0=ct\,,
\end{align}
where $\mathbf{v}$ is the non-relativistic velocity, $\mathbf{v}\equiv d\boldsymbol{x}/dt$. From the spatial components of the Grassmanian variables, the
conventional spin vector is defined as $S^i\equiv -\frac{i}{2} \epsilon^{ijk}\psi^j\psi^k$. Using $B^i = \frac{1}{2}\epsilon^{ijk} F^{jk}$ and $E^i = F^{0i}$,  
we can therefore express 
\begin{align}
-i\psi^0 F_{0i}\psi^i&=\frac{\boldsymbol{S}\cdot(\boldsymbol{\pi}\times\boldsymbol{E})}{c\pi^0}\,,\\
-\frac{i}{2}\psi^i F_{ij}\psi^j&=\boldsymbol{S}\cdot\boldsymbol{B} \, . 
\end{align}
Furthermore, in the non-relativistic limit, the electromagnetic ``Larmor" energy is small compared to the mass- we can therefore approximate\footnote{We note that due the Grassman nature of $X$ there is only one further non-zero term in this expansion $\propto X^2$. Due to the nilpotency of the Grassmanian variables this term is antisymmetric in four Lorentz indices and thus reminicent of the discussion in section \ref{sec:WLframework:imag}. We note however that in section \ref{sec:WLframework:imag}, the emergence of the anomaly was tied to the 
existence of Grassmannian zero modes and thereby resulted in the well known anomaly relation \Eq{eq:finalanomaly}. The order $X^2$  term here corresponds to a field configuration $\propto\mathbf{E}\cdot\mathbf{B}$; however it 
is not a sign of the presence of the anomaly and not related to the non-conservation of the axial current. See also \cite{Barducci:1982yw}, where such a term is seen in the equations of motion.}
\begin{align}
m_R=\sqrt{m^2+i\psi^\mu F_{\mu\nu}\psi^\nu}\approx m\left( 1+\frac{i}{2}\frac{\psi^\mu F_{\mu\nu}\psi^\nu}{m^2 c^2} \right) \equiv m(1+X),
\end{align}
where we introduced the abreviation
\begin{align}
X\equiv-\frac{\boldsymbol{S}\cdot(\boldsymbol{\pi}\times\boldsymbol{E})/(c\pi^0)+\boldsymbol{S}\cdot\boldsymbol{B}}{2\,m^2\,c^2}
\end{align}
The Lagrangian, which is defined by
\begin{align}
S=\int dt \;\mathcal{L}',
\end{align}
can be written as
\begin{align}
\mathcal{L}'&=-\frac{m_R c^2}{2\gamma}\left( 1+\frac {m^2}{m_R^2} \right)+\frac{i}{2}\left(\boldsymbol{\psi} \dot{\boldsymbol{\psi}}- \psi_0\dot{\psi}_0 \right) - A^0 +\frac{\boldsymbol{v}}{c}\cdot\boldsymbol{A}
+\frac{1}{m_R\gamma}\left( \frac{\boldsymbol{S}\cdot(\boldsymbol{\pi}\times\boldsymbol{E})}{c\pi^0}+\boldsymbol{S}\cdot\boldsymbol{B}\right)\nonumber\\
&=-\frac{mc^2}{2\gamma}\left( 1+X+\frac{1}{1+X}\right) +\frac{i}{2}\left(\boldsymbol{\psi} \dot{\boldsymbol{\psi}}- \psi_0\dot{\psi}_0 \right)- A^0 +\frac{\boldsymbol{v}}{c}\cdot\boldsymbol{A}-\frac{2\,mc^2}{\gamma}\frac{X}{1+X}.
\end{align}
The non-relativistic limit is found when $x\propto (\boldsymbol{v}/c)^2$ is small. Thus we expand the expression in terms of $X$ and $\boldsymbol{v}/c$ and keep only terms at most quadratic in the latter. This gives 
\begin{align}
\mathcal{L}'\approx-mc^2+\frac{1}{2}m \boldsymbol{v}^2+\frac{i}{2}\left(\boldsymbol{\psi} \dot{\boldsymbol{\psi}}- \psi_0\dot{\psi}_0 \right) + A^0 -\frac{\boldsymbol{v}}{c}\cdot\boldsymbol{A}
+\frac{ \boldsymbol{S}\cdot(\boldsymbol{\pi}\times\boldsymbol{E})}{mc\,\pi^0}+\frac{\boldsymbol{S}\cdot\boldsymbol{B}}{m}.
\end{align}
Since in this limit
\begin{align}
\pi^0\rightarrow p^0-\frac{A^0}{c}\,,\qquad\qquad{\rm and} \qquad\qquad
\boldsymbol{\pi}\rightarrow\boldsymbol{p}-\frac{\boldsymbol{A}}{c}\,,
\end{align}
we obtain our final form for the non-relativistic Lagrangian to be 
\begin{align}
\mathcal{L}_{NR}=-mc^2+\frac{1}{2}m \boldsymbol{v}^2+\frac{i}{2}\left(\boldsymbol{\psi} \dot{\boldsymbol{\psi}}- \psi_0\dot{\psi}_0 \right) - A^0 +\frac{\boldsymbol{v}}{c}\cdot\boldsymbol{A}
+ \frac{ \boldsymbol{S}\cdot(\left[ {\boldsymbol{v}/c}-{\boldsymbol{A}/(mc^2)}\right]\times\boldsymbol{E})}{mc}+\frac{\boldsymbol{S}\cdot\boldsymbol{B}}{m}\,.
\end{align}
Here $\psi_i$, $i=1,2,3$ are the dynamical spin degrees of freedom. Since $\psi_0$ is not dynamical, we shall drop it from now on. To obtain the corresponding non-relativistic Hamiltonian, we proceed just as we had done in the
Lorentz covariant case, by introducing a non-relativistic conjugate momentum
\begin{align}
p^i=\frac{\partial \mathcal{L}_{NR}}{\partial \dot{x}^i}=m\dot{x}^i+\frac{A^i}{c}+\frac{\epsilon^{ijk}E^j S^k}{mc^2} \,.
\end{align}
We can then compactly express the non-relativistic action as
\begin{align}\label{eq:NRaction}
S=\int d t\left( \boldsymbol{p}\cdot\dot{\boldsymbol{x}}+\frac{i}{2}\boldsymbol{\psi}\cdot\dot{\boldsymbol{\psi}}-H \right) \,,
\end{align}
where the non-relativistic Hamiltonian (in SI units) is
\begin{align}\label{eq:NRhamilt}
H\equiv mc^2+\frac{\left(\boldsymbol{p}-\frac{\boldsymbol{A}}{c}\right)^2}{2m}+A^0(x)
-\frac{ \boldsymbol{S}\cdot(\left[ {\boldsymbol{v}/c}-{\boldsymbol{A}/(mc^2)}\right]\times\boldsymbol{E})}{2mc}-\frac{\boldsymbol{B}\cdot\boldsymbol{S}}{m}\,.
\end{align}
This expression is of course the well known expression for the Hamiltonian for a fermion in an external electromagnetic field~\cite{Sakurai}: the penultimate term is the spin-orbit interaction energy from Thomas precession, while the last term is the Larmor interaction energy.

In the accompanying letter \cite{Mueller:2017lzw},  we showed in some detail that in an \textit{adiabatic approximation} the system
described by \Eq{eq:NRaction} and \Eq{eq:NRhamilt} contains a Berry phase with monopole form, also postulated in \cite{Son:2012wh,Stephanov:2012ki,Son:2012zy,Chen:2013iga}. In the next subsection, we will repeat part of our  derivation for the case of a massless particle in the presence of a large chemical potential. This is the case discussed for instance in \cite{Son:2012wh,Stephanov:2012ki} and several other works.

\subsection{Chemical Potential}\label{sec:chemicalpotential}
The limit that we derived in \Eq{eq:NRhamilt} is different from the one in \cite{Son:2012wh,Stephanov:2012ki}, as the latter contains an effective description for (massless) particles near the Fermi surface, which is well defined for large $\mu$. We will here explore how this limit appears in the world-line framework. 
As suggested by \Eq{eq:physicalsubspace}, a chemical potential can be introduced by adding a term to the Dirac operator equation
\begin{align}
\gamma_5\gamma^\nu\pi_\nu| \Phi\rangle=0\qquad \rightarrow \qquad \gamma_5(\gamma^\nu\pi_\nu +\mu\gamma^0 )| \Phi\rangle=0, 
\end{align}
The corresponding world-line expression is 
\begin{align}
\pi_\nu \psi^\nu=0\qquad \rightarrow \qquad \pi_\nu \psi^\nu + \mu \psi^0=0.
\end{align}
The mass-shell constraint is modified by the introduction of a chemical potential to read: 
\begin{align}
\pi^2 + i \psi^\alpha F_{\alpha\beta}\psi^\beta +\mu^2=0.
\end{align}

The world-line Lagrangian for massless fermions in the presence of a chemical potential is then\footnote{For simplicity, we have omitted the kinetic terms for $\psi_5$ and $\psi_6$.}
\begin{align}\label{eq:lagrangianmu}
\mathcal{L}(\mu)=\frac{\dot{x}^2}{2\mathcal{E}}-\frac{\mathcal{E}}{2}\mu^2 +\frac{i}{2}\psi_\alpha\dot{\psi}^\alpha+\dot{x}_\alpha A^\alpha - \frac{i\mathcal{E}}{2}\psi^\alpha F_{\alpha\beta}\psi^\beta 
-\frac{i}{2}\left(\frac{\dot{x}_\alpha\psi^\alpha}{\mathcal{E}}+\mu\psi^0\right)\chi\,,
\end{align}
which we emphasize is a relativistic expression. The path integral we have to evaluate is
\begin{align}\label{eq:pathintchemicalpotential}
W(\mu)=\int \frac{dT}{T}\int\mathcal{D}x\int\mathcal{D}\psi\,\exp{\big\{i\int\limits_0^T d\tau\,\mathcal{L}(\mu)   \big\} }.
\end{align}
As previously for \Eq{eq:consistenycond},  a consistency relation can be derived here as well.  In this case, we will proceed by performing the $T$ integration in \Eq{eq:pathintchemicalpotential} directly. The integral in \Eq{eq:pathintchemicalpotential} can be performed by the stationary phase method. Fixing $\mathcal{E}=2$, we obtain
\begin{align}
\mathcal{L}(\mu)= \int\limits_0^1 du\,\left\{\frac{\dot{x}^2}{4T} - \mu^2\left( 1 + \frac{i}{\mu^2}\psi^\alpha F_{\alpha\beta}\psi^\beta \right)T + \dot{x}_\alpha A^\alpha 
+ \frac{i}{2}\psi_\alpha\dot{\psi}^\alpha -\frac{i}{2}\left(\frac{\dot{x}_\alpha\psi^\alpha}{2}+\mu\psi^0\right)\chi\right\}.
\end{align}
We further rescale $T\rightarrow \int_0^1 du\, \mu^2( 1 + \frac{i}{\mu^2}\psi^\mu F_{\mu\nu}\psi^\nu)T\equiv m_\text{eff}^2T$ to obtain 
\begin{align}\label{eq:Wmu}
W(\mu)=\int \frac{dT}{T}e^{-iT}\int\mathcal{D}x\int\mathcal{D}\psi\,\exp{\left\{
i \frac{m_\text{eff}^2}{T}\int\limits_0^1 du\,\frac{\dot{x}^2}{4} + i\int\limits_0^1 du\,\left\{\dot{x}_\mu A^\mu 
+ \frac{i}{2}\psi_\mu\dot{\psi}^\mu -\frac{i}{2}\left(\frac{\dot{x}_\alpha\psi^\alpha}{2}+\mu\psi^0\right)\chi \right\} 
\right\} }.
\end{align}
For large chemical potential the intergral is dominated by the first term in the exponent. Therefore, using the stationary phase method,
the $T$ integral can be performed around the stationary point $T_0=m_\text{eff}\sqrt{- \int_0^1 du\,\frac{\dot{x}^2}{4}}$. The result is
\begin{align}
W(\mu)\approx \int\mathcal{D}x&\int\mathcal{D}\psi\,\sqrt{\frac{i\pi}{2m_\text{eff}}}\left(-\int_0^1 du\, \dot{x}^2 \right)^{-\frac{1}{4}}\nonumber\\\times &\exp{\left\{ -im_\text{eff} \sqrt{-\int_0^1 du\,  \dot{x}^2} 
+i \int\limits_0^1 du\,\left(\dot{x}_\alpha A^\alpha + \frac{i}{2}\psi_\alpha\dot{\psi}^\alpha -\frac{i}{2}\left[\frac{\dot{x}_\alpha\psi^\alpha}{2}+\mu\psi^0\right]\chi \right)\right\} }.
\end{align}
For a large chemical potential, we can Taylor expand
\begin{align}
m_\text{eff}  \approx \mu\left( 1 + \frac{i}{2\mu^2}\int_0^1 du\,\psi^\alpha F_{\alpha\beta}\psi^\beta \right),
\end{align}
so that we finally have--using the abreviation $\mathcal{\bar{N}}\equiv\sqrt{\frac{i\pi}{2m_\text{eff}}}\left(-\int_0^1 du\, \dot{x}^2 \right)^{-\frac{1}{4}}$,
\begin{align}\label{eq:pathintegral_chemicalpot}
W(\mu)\approx \int\mathcal{D}x\int\mathcal{D}\psi\,\mathcal{\bar{N}}\,
\exp{ \left\{ 
-i \mu\,\sqrt{ -\int_0^1 du\,\dot{x}^2} 
+i \int\limits_0^1 du\,\left(\frac{-i}{2\mu}\psi^\alpha F_{\alpha\beta}\psi^\beta+ \dot{x}_\alpha A^\alpha + \frac{i}{2}\psi_\alpha\dot{\psi}^\alpha  \right)
\right\} }.
\end{align}
This effective action describes excitations near the fermion surface for a massless theory with a large chemical potential. In obtaining this form for the action, in analogy with the previous section, we chose $\chi=0$. 
\Eq{eq:pathintegral_chemicalpot}  can be directly compared with the result in section \ref{sec:NRadiabatic}: as might have been anticipated, the role of the mass parameter is effectively taken over by the chemical potential. 
The non-relativistic limit is thus identical upon this identification, as is the adiabatic limit in \cite{Mueller:2017lzw}. It was shown there how a Berry monopole is found when level crossings between spin states are suppressed.

A closer look at the individual terms in the action of $W(\mu)$ illustrates these points nicely. While the first square root term is the conventional kinetic term for a particle with effective mass $\mu$, 
the second term is a Larmor-interaction energy, with the effective mass  $\mu$. For large chemical potentials, excitations around the Fermi surface behave non-relativistically. Further, the adiabatic limit corresponds 
to $\psi^\alpha F_{\alpha\beta}\psi^\beta/\mu\approx 0$. The effective description of \cite{Son:2012wh,Stephanov:2012ki} is thus straightforwardly understood by taking the appropriate limits in the world-line framework. 

The aforementioned non-relativistic and the adiabatic approximation may not be applicable to ultra-relativistic heavy-ion collisions. Instead, the general Lorentz-covariant world-line
framework, which we have established in Eqs.~(\ref{eq:eomx1}-\ref{eq:eompsi5}) is ideally suited for the description of the anomalous transport of axial charges in the hot fireball created in a heavy-ion collision.

\section{Conclusions}\label{sec:conclusions}
In this manuscript, and in an accompanying letter \cite{Mueller:2017lzw}, we developed a world line framework in quantum field theory to construct a Lorentz-covariant chiral kinetic theory for fermions.  In the first part of the paper, we 
obtained a world-line path integral representation of the (Euclidean) fermion determinant in the background of vector and axial-vector gauge fields. This was achieved by using a heat-kernel representation of 
the (infinite-dimensional) operator logarithm. We exploited a fermionic coherent state formalism whereby spin is not treated as part of a wave function but rather as an independent degree of freedom in the path 
integral. This powerful construction can be extended to include other internal degrees of freedom such as color.

We then investigated how the axial anomaly arises in world line quantization. As is well known \cite{AlvarezGaume:1983ig}, the axial anomaly is related to the phase of the fermion determinant, which is ill-defined for fermions 
in a complex representation. The fermion effective action is thus understood to contain both a real as well as an imaginary part, the latter being related to the violation of chiral symmetry. Using a path integral construction due to  D'Hoker and Gagn\'e~\cite{D'Hoker:1995ax,D'Hoker:1995bj},  we obtained a representation of the real part of the effective action in terms of a Grassmanian path integral over spinning variables. Remarkably, there is a 
very similar path integral representation  for the imaginary part, wherein an integral over a regulating parameter represents the loss of chiral symmetry. This path integral representation includes an operator insertion, which in 
this framework is responsible for the fermion zero modes in the spectrum of the theory. Following the discussion by Alvarez-Gaume and Witten~\cite{AlvarezGaume:1983ig}, we demonstrated in our framework how these modes are 
responsible for the axial anomaly. In particular, we employed a variational method to obtain a non-perturbative expression for the axial-vector current in first-quantization and thence derived the anomaly equation. 
The emergence of the axial anomaly in first quantization  crucially depends on a hidden supersymmetry between bosonic and fermionic degrees of freedom induced by periodic boundary conditions for the fermion variables on the closed world-line.

Motivated by our findings in section \ref{sec:ckt}, we derived the pseudo-classical kinetic limit of the world-line effective action. Continuing our prior discussion from Euclidean to Minkowski metric, we established that 
the Liouville dynamics of spinning particles arises from the real part (in the original Euclidean formulation) of the fermion determinant alone.  This contribution to the kinetic dynamics is independent of those arising 
from the piece in the path integral containing the fermionic zero modes that are responsible for the anomaly. However in a chiral kinetic theory, anomalous contributions to the dynamics, in a covariant formulation,  
will be manifest through the axial vector current.

A part of the impetus of our work was to understand the origins of the Berry term in kinetic descriptions from first principles in quantum field theory and to establish thereby its relation, if any, to the chiral anomaly. 
In our accompanying letter \cite{Mueller:2017lzw}, we showed how such a term arises from the world-line action for massive spinning particles in external background gauge fields. We demonstrated explicitly that we 
needed to take the non-relativistic limit of large masses, as well as an adiabatic limit wherein the Larmor interaction energy of the spinning particles was much smaller than the rest energy. In this paper, we addressed
the problem in the case where the spinning particles are massless but the system possesses a large chemical potential. This is the case for quasi-particle excitations near the Fermi surface in a number of condensed matter systems. We showed explicitly in the world-line framework that the chemical potential replaces the role of the mass and the rest energy in a manner that is exactly the same as was the case for massive spinning particles. An anaologous non-relativistic and adiabatic limit for these excitations can therefore be taken, and it can be similarly be demonstrated how the Berry term arises upon taking these limits. 

This exercise also suggests that away from the adiabatic non-relativistic limit,  the Berry phase is not robust and 
its effects are implicit in the relativistic dynamics of spinning particles. As such, we have arrived at the same conclusion as the previous observation by Fujikawa and collaborators~\cite{Deguchi:2005pc,Fujikawa:2005tv,Fujikawa:2005cn}. In contrast to the Berry phase, the effects of the anomaly are robust and manifest in a relativistic kinetic description. More generally, the semi-classical world-line construction we obtained here can be incorporated in a real-time Schwinger-Keldysh framework to describe the evolution of a chiral current in a gauge field background. A similar such construction was performed in \cite{JalilianMarian:1999xt} for spinless colored particles. It was shown in that case how one recovers in the world-line framework the non-Abelian Boltzmann Langevin ``B\"{o}deker kinetic theory"~\cite{Bodeker:1998hm,Bodeker:1999ey,Litim:1999id} of hot QCD. This framework can be extended to construct an ``anomalous B\"{o}deker theory"  which can then be matched to classical-statistical simulations at early times in heavy-ion collisions and to anomalous hydrodynamics at late times. This work is in progress~\cite{follow-up}. We note that there has been a recent discussion of the anomalous B\"{o}deker kinetic theory in the literature~\cite{Akamatsu:2014yza,Akamatsu:2015kau} in a different approach and it will be useful in future to compare and contrast results in the two approaches.  

The framework presented here is not only applicable in the QCD framework of heavy-ion collisions but is potentially applicable to a number of many-body contexts where 
topology is important and the dynamics is relativistic. One such example is that of the transport of chiral
fermions in an astrophysical situations~\cite{Charbonneau:2009ax,Akamatsu:2013pjd,Dvornikov:2016gdo,Kaplan:2016drz,Akamatsu:2013pjd}. In this context,  our framework provides a first principles perspective that can be used to address situations where masses and chemical potentials are not large and non-relativistic and 
adiabatic assumptions are no longer valid. Another intriguing possibility is to apply this framework to helicity evolution in QCD at small $x$~\cite{Kovchegov:2016weo}. In QCD at small $x$, semi-classical concepts provide fertile 
ground\cite{McLerran:1993ni,McLerran:1993ka}; a semi-classical world-line description was previously employed~\cite{JalilianMarian:2000ad} to derive the 
well known BFKL equation for unpolarized parton distributions~\cite{Kuraev:1977fs,Balitsky:1978ic}. The world-line construction developed here for spinning particles therefore shows great promise for a wide of many-body problems and will be pursued in future work. 

\section*{Acknowledgments}
RV thanks the Institut f\"{u}r Theoretische Physik, Heidelberg for their kind hospitality and the Excellence Initiative of Heidelberg University for a Guest Professorship during the period when this work was initiated. We thank Juergen Berges, Jan Pawlowski, and Michael Schmidt for encouraging this effort. We thank Fiorenzo Bastianelli for many helpful discussions and for sharing his deep insights into the world-line formulation of 
quantum field theory. We thank Cristina Manuel and Naoki Yamamoto for valuable comments on the accompanying letter.
RV would also like to thank the attendees of a seminar on this work at Stony Brook for their helpful comments; in particular, he would like to thank Dima Kharzeev, Ho-Ung Yee, Yi Yin and Ismail Zahed. NM further thanks Valentin Kasper for helpful discussions.

NM acknowledges support by the Studienstiftung des Deutschen Volkes and by the DFG Collaborative Research Centre SFB 1225 (ISOQUANT). This material is partially based upon work supported by the U.S. Department of Energy,
Office of Science, Office of Nuclear Physics, under contract No. DE- SC0012704, and within the framework of the Beam Energy Scan Theory (BEST) Topical Collaboration.
%
%
%
%
\begin{appendix}\label{sec:app}
\section{Details of the calculation of the imaginary part of the effective action}\label{sec:app:anomalydetails}
In this Appendix, as promised, we will show that the second term in the world-line insertion,  does not contribute to the non-conservation of the axial vector current. Writing out the relevant expression, 
\begin{align}
\partial_\mu\text{Tr}&\left( \Gamma_7 \frac{\delta\Lambda^{(2)}}{\delta B_\mu(y)} e^{-\frac{\mathcal{E}}{2}T\tilde{\Sigma}^2} \right)=-\int\left(\prod\limits_{l=0}^Nd^4x^ld^3\theta^ld^3\bar{\theta}^l \right)\left(\prod\limits_{l=1}^N\frac{d^4p^l}{(2\pi)^2} \right)\langle \theta^0| 
[\Gamma_\mu,\Gamma_\nu]\Gamma_5\Gamma_6| \theta^N\rangle
\Big[\left( \frac{\partial^2}{\partial y_\mu \partial x^0_\mu}\delta(x^0-y)\right) \delta(x^0-x^N)\nonumber\\
&+2\left( \frac{\partial}{\partial x_\nu}\delta(x^0-x^N) \right) \left(\frac{\partial}{\partial y_\mu} \delta(\bar{x}^0-y)\right) \Big]
\exp\Big\{
-\Delta\sum\limits_{k=1}^N\big[%
-ip_\alpha^k\frac{(x_\alpha^k-x_\alpha^{k-1})}{\Delta} +\frac{\mathcal{E}}{2}\left( p_\alpha^k-A_\alpha(\bar{x}^k) \right)^2\nonumber\\
&\qquad\qquad\qquad\qquad\qquad\qquad\qquad\qquad\qquad\qquad\qquad\qquad-\frac{(\theta^k_r-\theta^{k-1}_r)}{\Delta}\bar{\theta}^k_r+\frac{i\mathcal{E}}{2}\psi^k_\alpha\psi^{k-1}_\beta F_{\alpha\beta}(\bar{x}^k)
\big]
\Big\}
\nonumber\\=&-\int\left(\prod\limits_{l=0}^{N-1}d^4x^l\right)\left(\prod\limits_{l=0}^Nd^3\theta^ld^3\bar{\theta}^l \right)\left(\prod\limits_{l=1}^N\frac{d^4p^l}{(2\pi)^2} \right)\langle \theta^0| 
[\Gamma_\mu,\Gamma_\nu]\Gamma_5\Gamma_6| \theta^N\rangle \nonumber\\&\times\Big[-\frac{\partial^2}{\partial x^0_\mu \partial x^0_\nu} \Big]\exp\Big\{
-\Delta\sum\limits_{k=1}^N\big[%
-ip_\alpha^k\frac{(x_\alpha^k-x_\alpha^{k-1})}{\Delta} +\frac{\mathcal{E}}{2}\left( p_\alpha^k-A_\alpha(\bar{x}^k) \right)^2
-\frac{(\theta^k_r-\theta^{k-1}_r)}{\Delta}\bar{\theta}^k_r+\frac{i\mathcal{E}}{2}\psi^k_\alpha\psi^{k-1}_\beta F_{\alpha\beta}(\bar{x}^k)
\big]
\Big\}
\end{align}
In the final expression above,  we observe that while the expression containing the commutator of Gamma matrices is anti-symmetric under the exchange of $\mu$ and $\nu$, the derivative of the exponent is clearly symmetric under this exchange. Therefore
\begin{align}
\partial_\mu\text{Tr}&\left( \Gamma_7 \frac{\delta\Lambda^{(2)}}{\delta B_\mu(y)} e^{-\frac{\mathcal{E}}{2}T\tilde{\Sigma}^2} \right)=0 \,,
\end{align}
which completes our proof of the statement following \Eq{eq:wlinsertiondef} in the main text of the paper. 
%
%
\section{Supersymmetry and gauge freedom of the relativistic spinning particle}\label{app:susy}
The Lorentz-covariant formulation of the spinning particle action given by \Eq{eq:lagrangian:RealTime} posses two important symmetries respected by the world-line path integral.
Firstly, the physical content of the theory is invariant under reparametrizations of the world line parameter $\tau$,
\begin{align}
\tau\rightarrow\tau'=f(\tau)
\end{align}
This gauge symmetry corresponds to the mass-shell constraint or ``charge"
\begin{align}\label{eq:constr1}
\mathcal{H}\equiv\frac{1}{2}\left(\pi_\mu\pi^\mu+m^2+i\psi^\mu F_{\mu\nu} \psi^\nu\right)\,,
\end{align}
(with $\pi^\mu$ defined as in \Eq{eq:conjugate}) which upon quantization is a constraint on the physical states in the Hilbert space--equivalent to the Klein-Gordon equation. It is also closely connected to another invariance of the action in terms
of proper time dependent quantum mechanical supersymmetric transformations. Assuming $\eta(\tau)$ to be an anti-commuting parameter, these  supersymmetric transformations are 
\begin{align}
\psi_\mu&\rightarrow\psi_\mu+\frac{\dot{x}_\mu}{\sqrt{-\dot{x}^2}}\,\eta \,,\nonumber\\
\psi_5&\rightarrow\psi_5+\eta\,,\nonumber\\
x_\mu&\rightarrow x_\mu+i\frac{\psi_\mu\eta}{m}\,.
\end{align}
These transformations correspond to the supersymmetric charge, 
\begin{align}\label{eq:constr2}
\mathcal{Q}\equiv\pi_\mu\psi^\mu+m\psi_5\,.
\end{align}
This charge, along with the constraints \Eq{eq:constr1} and \Eq{eq:constr2}, generates an $N=1$ SUSY algebra,
\begin{align}
\left\{\mathcal{Q},\mathcal{Q}\right\}=-2i\,\mathcal{H}.
\end{align}
In generating this algebra, one employs the fundamental Possion brackets:
\begin{align}
\left\{x^\mu, p_\nu \right\}&=\delta^\mu_{\;\nu}\,,\\
\left\{\psi^\mu, \psi_\nu \right\}&=-i\delta^\mu_{\;\nu}\,,\\
\left\{\psi_5, \psi_5 \right\}&=-i\,,\\
\left\{\psi^\mu, \psi_5 \right\}&=0\,.
\end{align}
We refer the reader to \cite{AlvarezGaume:1983ig,FiorenzoBookNew} for more details on the use of SUSY models in the context of path integrals and index theorems, as they are used, most prominently, in gravity. 
A discussion of a covariant of a covariant fixing of the gauge freedom (reparamentrization invariance under $\tau\rightarrow\tau'$) in terms of a BRST construction can be found in \cite{FiorenzoBookNew} and gives a nice illustration of the structure of the world-line path integral. 
These techniques will be particularly helpful in implementing the phase space constraints satisfied by the relativistic dynamics of spinning and colored particles. 
\section{Internal Symmetries}\label{app:internalsymmetries}
Internal symmetries, such as color, can be represented via Grassmaniann path integrals in the same manner as we have done for the spin degrees of freedom. These were
discussed in \cite{Berezin:1976eg,Balachandran:1976ya,Balachandran:1977ub,Barducci:1982yw,Brink:1976uf} and their path integral formulation
was worked out in \cite{D'Hoker:1995ax,D'Hoker:1995bj}. The essential elements are anti-commuting color degrees of freedom that combine to give the color charges, which in classical representations satisfy the Wong equations~\cite{Wong:1970fu}. 
It was shown in \cite{D'Hoker:1995ax,D'Hoker:1995bj} that path ordered exponentials of the form 
\begin{align}
\text{tr}\;\mathcal{P}e^{-\int\limits_0^Td\tau\;\mathcal{L}(\tau)}, 
\end{align}
where $\mathcal{L}(\tau)$ is a $N\times N$ Hermitian traceless matrix, can be written as
\begin{align}
\int\mathcal{D}
\lambda^\dagger\mathcal{D}\lambda\,\mathcal{J}(\lambda^\dagger\lambda)
\,\exp{\left\{-\int_0^T d\tau\,\left( \frac{\dot{x}^2}{2\mathcal{E}}+\frac{1}{2}\psi_a\dot{\psi}_a + \lambda^\dagger\dot{\lambda} - \lambda^\dagger \mathcal{L}_\text{int}\lambda \right) \right\}},
\end{align}
where $\mathcal{L}_\text{int}$ is the interaction part of the Lagrangian and $\mathcal{J}(\lambda^\dagger\lambda) = (\frac{\pi}{T})^{N} \sum_\phi \exp[ i\phi (\lambda^\dagger \lambda + N/2 -1)]$. If the matrix structure of $\mathcal{L}$ is that of fermions in the fundamental representation of $SU(N_c)$, then simply $N=N_c$.
In a similar fashion, an insertion $\omega$ into the trace gives
\begin{align}
\text{tr}\mathcal{P}\,\omega\;e^{-\int\limits_0^Td\tau\;\mathcal{L}(\tau)}=\int\mathcal{D}
\lambda^\dagger\mathcal{D}\lambda\;\mathcal{J}(\lambda^\dagger\lambda)\,\{ \lambda^\dagger \omega \lambda\}\;e^{-\int\limits_0^T d\tau\;\left( \frac{\dot{x}^2}{2\mathcal{E}}+\frac{1}{2}\psi_a\dot{\psi}_a + \lambda^\dagger\dot{\lambda} - \lambda^\dagger \mathcal{L}_\text{int}\lambda \right) }\, .
\end{align}
 It can be shown that defining the world-line path integral for colored, albeit spinless, particles reproduces Wong's equations in the pseudo-classicial limit \cite{JalilianMarian:1999xt}. 
The equations of motion for spinning colored particles were already written down 40 years ago in \cite{Barducci:1976xq}. 
\section{Consistent vs. Covariant Anomalies}\label{app:anomalies}
It has long been known that the definition of axial-vector currents is ambiguous in some cases, allowing for two anomaly types, termed consistent
and covariant respectively. It was pointed out \cite{Bardeen:1984pm} that this difference arises when one derives the non-singlet
anomaly either from the variation of an effective action (which yields the consistent anomaly) or from Fujikawa's method via variation of the measure (which gives the covariant anomaly). The first type was called the consistent anomaly, as it fullfils the Wess-Zumino consistency conditions thereby predicting the correct anomalous physics of effective hadronic theories. The second type is obtained from the first type by adding a local counterterm, which makes the non-singlet anomaly transform covariantly under group transformations.

For the singlet anomaly, and in QED, this issue is much simpler, as in this case one has manifestly gauge \textit{invariant} expressions for both vector and axial-vector currents. Therefore the possibility that a current is not covariant
never arises. However as was discussed by Bardeen \cite{Bardeen:1969md}, care has to be taken when deriving currents, when there are both non-zero vector- as well as axial-vector fields. In this case, there is an ambiguity
whether the anomaly should be contained in the vector- or the axial-vector-currents (or even both). Physics dictates that the vector current is related to the baryon number and so it better be conserved. Hence by the introduction of local Bardeen-counterterms 
this physicality condition can be enforced. We note however that if there are no physical axial-vector gauge fields present, as it is in our case, this ambiguity does not exist. The vector current is conserved by 
construction and hence the only possible form of the anomaly is given by \Eq{eq:finalanomaly}.
\section{Saddle Point Expansion in the World Line framework and Gauge Invariance}\label{sec:saddlepointdef}
We will discuss here two different appraoches to the fixing of the gauge symmetry determining $\mathcal{E}$. Our derivation is based on the fact that $\mathcal{E}$ is related to the reparametrization invariance of the proper time
\begin{align}
\tau\rightarrow \tau'=f(\tau),
\end{align}
where $f$ is an arbitrary continuous function. For the sake of simplicity, we will neglect here spin dependent pieces of our action and write down the world-line path integral for a scalar particle. We will then subsequently generalize the discussion to particles with spin. 
The world-line path integral for the spinless case is 
\begin{align}\label{eq:worldlinescalar}
W_\text{scalar}&=\int\limits_0^\infty\frac{dT}{T}\int\mathcal{D}x\;\exp{\left(i\int_0^T d\tau\;\Big[\frac{\dot{{x}}^2}{2\mathcal{E}}+\dot{x}_\mu A^\mu(x)-\frac{\mathcal{E}}{2}m^2\Big] \right)}\nonumber\\
&=\int\limits_0^\infty\frac{dT}{T}\int\mathcal{D}x\;\exp{\left(i\int_0^1 du\;\Big[\frac{(dx/du)^2}{2\mathcal{E}T}+\frac{dx_\mu}{du} A^\mu(x)-\frac{\mathcal{E}T}{2}m^2\Big] \right)},
\end{align}
where in the second line we have replaced $u=\tau/T$. From \Eq{eq:worldlinescalar} it is clear that $T$ and $\mathcal{E}$ are not independent. Setting $\mathcal{E}$ to a constant value does not affect the result of the $T$
integration. Therefore we can simply set $\mathcal{E}=2$ and rescale $m^2 T\rightarrow T$. The path integral is then given as
\begin{align}\label{eq:worldlinescalar1}
W_\text{scalar}=\int\limits_0^\infty\frac{dT}{T}\int\mathcal{D}x\;\exp{\left(i\int_0^1 du\;\Big[m^2\frac{(dx/du)^2}{4T}+\frac{dx_\mu}{du} A^\mu(x)-T\Big] \right)},
\end{align}
The $T$ integration can now either be performed explicitly (see \cite{Dunne:2005sx,Dunne:2006st}) or via the method of stationary phase around the expansion point
\begin{align}\label{eq:expansionpoint}
T_0=\frac{m}{2}\left(-\int_0^1du\;\left[\frac{dx_\mu}{du}\right]^2\right)^\frac{1}{2}. 
\end{align}
We obtain
\begin{align}
W_\text{scalar}\approx\sqrt{\frac{i\pi}{2m}}\int \mathcal{D}x\;\left(-\int_0^1du\;\left[\frac{dx_\mu}{du}\right]^2\right)^{-\frac{1}{4}}\exp{i\left\{m\left(-\int_0^1du\;\left[\frac{dx_\mu}{du}\right]^2\right)^\frac{1}{2}
+\int_0^1du\;
\frac{dx_\mu}{du} A^\mu(x) \right\}}\,.
\end{align}
We now derive the equations of motion from requiring the invariance of this action under variation. The result is
\begin{align}
\left(-\int_0^1 du\;\dot{x}^2\right)^{-\frac{1}{2}}\; m\ddot{x}_\mu=\dot{x}_\nu F^{\mu\nu}.
\end{align}
 We can write this, defining $z=\sqrt{-\dot{x}^2}$, as\footnote{Multiplying this equation through by $\dot{x}^\mu$, one observes that $\dot{x}^2={\rm constant}$.}
\begin{align}\label{eq:eom1scalar}
\frac{ m\ddot{x}_\mu}{z}=\dot{x}_\nu F^{\mu\nu} \, .
\end{align}

One can alternately start from the Lagrangian 
in \Eq{eq:worldlinescalar}. Instead of fixing $\mathcal{E}$ and leaving the $T$ integral explicit, we can work with the single particle action 
\begin{align}\label{eq:actionconstitentnaive}
S=\int_0^T d\tau\;\Big[\frac{\dot{{x}}^2}{2\mathcal{E}}+\dot{x}_\mu A^\mu(x)-\frac{\mathcal{E}}{2}m^2\Big] 
\end{align}
directly. Since $\mathcal{E}$ is kept explicit, there are two variations to perform -- one for $\mathcal{E}$ and one with respect to $x$. Variation with respect to $\mathcal{E}$
gives
\begin{align}
\frac{\dot{x}^2}{\mathcal{E}^2}-m^2=0\,.
\end{align}
Solving this equation for $\mathcal{E}$, one obtains the consistency relation
\begin{align}\label{eq:consistency}
\mathcal{E}=\frac{\sqrt{-\dot{x}^2}}{m}=\frac{z}{m}\,.
\end{align}
 Note that \Eq{eq:consistency} does not fix the gauge, as $z$ has yet to be determined.  It rather is an equation
that allows us to implement the constraint in the action directly. Plugging \Eq{eq:consistency} into \Eq{eq:actionconstitentnaive} eliminates the dependence on the einbein parameter and yields
\begin{align}\label{eq:scalarexp1}
S=\int_0^T d\tau\;\Big[mz +\dot{x}_\mu A^\mu(x) \Big]\,,
\end{align}
from which the equations of motion follow directly. Not surprisingly, they coincide with \Eq{eq:eom1scalar}. This derivation shows that \Eq{eq:actionconstitentnaive} can be interpreted as 
a single-particle action, under the premise that all constraints are implemented correctly and the consistency condition \Eq{eq:consistency} is fulfilled. The latter is satisfied if 
the einbein $\mathcal{E}$ is treated as a variational parameter. This equivalence generalizes easily to the case of spinning particles as discussed in the main text.

\end{appendix}


\begin{thebibliography}{90}

\bibitem{Mueller:2017lzw} 
  N.~Mueller and R.~Venugopalan,
  arXiv:1701.03331 [hep-ph].
  
\bibitem{Klinkhamer:1984di} 
  F.~R.~Klinkhamer and N.~S.~Manton,
  Phys.\ Rev.\ D {\bf 30}, 2212 (1984).
  doi:10.1103/PhysRevD.30.2212
  
\bibitem{Dashen:1974ck} 
  R.~F.~Dashen, B.~Hasslacher and A.~Neveu,
  Phys.\ Rev.\ D {\bf 10}, 4138 (1974).
  doi:10.1103/PhysRevD.10.4138
  
\bibitem{Soni:1980ps} 
  V.~Soni,
  Phys.\ Lett.\  {\bf 93B}, 101 (1980).
  doi:10.1016/0370-2693(80)90104-5
  
\bibitem{Boguta:1983xs} 
  J.~Boguta,
  Phys.\ Rev.\ Lett.\  {\bf 50}, 148 (1983).
  doi:10.1103/PhysRevLett.50.148
  
\bibitem{Forgacs:1983yu} 
  P.~Forgacs and Z.~Horvath,
  Phys.\ Lett.\  {\bf 138B}, 397 (1984).
  doi:10.1016/0370-2693(84)91926-9



\bibitem{Sakharov:1967dj} 
  A.~D.~Sakharov,
  Pisma Zh.\ Eksp.\ Teor.\ Fiz.\  {\bf 5}, 32 (1967)
  [JETP Lett.\  {\bf 5}, 24 (1967)]
  [Sov.\ Phys.\ Usp.\  {\bf 34}, 392 (1991)]
  [Usp.\ Fiz.\ Nauk {\bf 161}, 61 (1991)].
  doi:10.1070/PU1991v034n05ABEH002497

\bibitem{Riotto:1999yt}
  A.~Riotto and M.~Trodden,
  Ann.\ Rev.\ Nucl.\ Part.\ Sci.\  {\bf 49} (1999) 35
  doi:10.1146/annurev.nucl.49.1.35
  [hep-ph/9901362].
  
\bibitem{Cohen:1993nk} 
  A.~G.~Cohen, D.~B.~Kaplan and A.~E.~Nelson,
  Ann.\ Rev.\ Nucl.\ Part.\ Sci.\  {\bf 43}, 27 (1993)
  doi:10.1146/annurev.ns.43.120193.000331
  [hep-ph/9302210].
 
\bibitem{Rubakov:1996vz} 
  V.~A.~Rubakov and M.~E.~Shaposhnikov,
  Usp.\ Fiz.\ Nauk {\bf 166}, 493 (1996)
  [Phys.\ Usp.\  {\bf 39}, 461 (1996)]
  doi:10.1070/PU1996v039n05ABEH000145
  [hep-ph/9603208].
  
\bibitem{cond-matter_review}
  M.~Gradhand, D.~V.~Fedorov , F.~Pientka, P.~Zahn, I.~Mertig and B.~L.~Gy\"orffy 2012
  J. Phys. Condens. Matter 24 (2012) 213202
  
\bibitem{Kharzeev:2007jp} 
  D.~E.~Kharzeev, L.~D.~McLerran and H.~J.~Warringa,
  Nucl.\ Phys.\ A {\bf 803}, 227 (2008)
  doi:10.1016/j.nuclphysa.2008.02.298
  [arXiv:0711.0950 [hep-ph]].
  
\bibitem{Fukushima:2008xe} 
  K.~Fukushima, D.~E.~Kharzeev and H.~J.~Warringa,
  Phys.\ Rev.\ D {\bf 78}, 074033 (2008)
  doi:10.1103/PhysRevD.78.074033
  [arXiv:0808.3382 [hep-ph]].

\bibitem{Kharzeev:2015znc} 
  D.~E.~Kharzeev, J.~Liao, S.~A.~Voloshin and G.~Wang,
  Prog.\ Part.\ Nucl.\ Phys.\  {\bf 88}, 1 (2016)
  doi:10.1016/j.ppnp.2016.01.001
  [arXiv:1511.04050 [hep-ph]].
  
  
\bibitem{Li:2014bha} 
  Q.~Li {\it et al.},
  Nature Phys.\  {\bf 12}, 550 (2016)
  doi:10.1038/nphys3648
  [arXiv:1412.6543 [cond-mat.str-el]].
  
\bibitem{Skokov:2016yrj} 
  V.~Skokov, P.~Sorensen, V.~Koch, S.~Schlichting, J.~Thomas, S.~Voloshin, G.~Wang and H.~U.~Yee,
  arXiv:1608.00982 [nucl-th].
  

  
\bibitem{Skokov:2009qp} 
  V.~Skokov, A.~Y.~Illarionov and V.~Toneev,
  Int.\ J.\ Mod.\ Phys.\ A {\bf 24}, 5925 (2009)
  doi:10.1142/S0217751X09047570
  [arXiv:0907.1396 [nucl-th]].
  
\bibitem{Deng:2012pc} 
  W.~T.~Deng and X.~G.~Huang,
  Phys.\ Rev.\ C {\bf 85}, 044907 (2012)
  doi:10.1103/PhysRevC.85.044907
  [arXiv:1201.5108 [nucl-th]].
  

  
\bibitem{Mace:2016svc} 
  M.~Mace, S.~Schlichting and R.~Venugopalan,
  Phys.\ Rev.\ D {\bf 93}, no. 7, 074036 (2016)
  doi:10.1103/PhysRevD.93.074036
  [arXiv:1601.07342 [hep-ph]].
  
\bibitem{Moore:2010jd} 
  G.~D.~Moore and M.~Tassler,
  JHEP {\bf 1102}, 105 (2011)
  doi:10.1007/JHEP02(2011)105
  [arXiv:1011.1167 [hep-ph]].

  
\bibitem{Berges:2017eom} 
  J.~Berges, K.~Reygers, N.~Tanji and R.~Venugopalan,
  arXiv:1701.05064 [nucl-th].

\bibitem{Tanji:2016dka} 
  N.~Tanji, N.~Mueller and J.~Berges,
  Phys.\ Rev.\ D {\bf 93}, no. 7, 074507 (2016)
  doi:10.1103/PhysRevD.93.074507
  [arXiv:1603.03331 [hep-ph]].
  
\bibitem{Mueller:2016ven} 
  N.~Mueller, S.~Schlichting and S.~Sharma,
  Phys.\ Rev.\ Lett.\  {\bf 117}, no. 14, 142301 (2016)
  doi:10.1103/PhysRevLett.117.142301
  [arXiv:1606.00342 [hep-ph]].
  
\bibitem{Mace:2016shq} 
  M.~Mace, N.~Mueller, S.~Schlichting and S.~Sharma,
  arXiv:1612.02477 [hep-lat].
  
  
\bibitem{Berges:2013eia} 
  J.~Berges, K.~Boguslavski, S.~Schlichting and R.~Venugopalan,
  Phys.\ Rev.\ D {\bf 89}, no. 7, 074011 (2014)
  doi:10.1103/PhysRevD.89.074011
  [arXiv:1303.5650 [hep-ph]].
  
\bibitem{Kurkela:2015qoa} 
  A.~Kurkela and Y.~Zhu,
  Phys.\ Rev.\ Lett.\  {\bf 115}, no. 18, 182301 (2015)
  doi:10.1103/PhysRevLett.115.182301
  [arXiv:1506.06647 [hep-ph]].
  
\bibitem{Son:2009tf} 
  D.~T.~Son and P.~Surowka,
  Phys.\ Rev.\ Lett.\  {\bf 103}, 191601 (2009)
  doi:10.1103/PhysRevLett.103.191601
  [arXiv:0906.5044 [hep-th]].
  
\bibitem{Gursoy:2014aka} 
  U.~Gursoy, D.~Kharzeev and K.~Rajagopal,
  Phys.\ Rev.\ C {\bf 89}, no. 5, 054905 (2014)
  doi:10.1103/PhysRevC.89.054905
  [arXiv:1401.3805 [hep-ph]].
  
\bibitem{Hongo:2013cqa} 
  M.~Hongo, Y.~Hirono and T.~Hirano,
  arXiv:1309.2823 [nucl-th].

\bibitem{Hirono:2014oda} 
  Y.~Hirono, T.~Hirano and D.~E.~Kharzeev,
  arXiv:1412.0311 [hep-ph].
  
\bibitem{Yin:2015fca} 
  Y.~Yin and J.~Liao,
  Phys.\ Lett.\ B {\bf 756}, 42 (2016)
  doi:10.1016/j.physletb.2016.02.065
  [arXiv:1504.06906 [nucl-th]].
  

\bibitem{Son:2012wh} 
  D.~T.~Son and N.~Yamamoto,
  Phys.\ Rev.\ Lett.\  {\bf 109}, 181602 (2012)
  doi:10.1103/PhysRevLett.109.181602
  [arXiv:1203.2697 [cond-mat.mes-hall]].
  
\bibitem{Stephanov:2012ki} 
  M.~A.~Stephanov and Y.~Yin,
  Phys.\ Rev.\ Lett.\  {\bf 109}, 162001 (2012)
  doi:10.1103/PhysRevLett.109.162001
  [arXiv:1207.0747 [hep-th]].
  
\bibitem{Son:2012zy} 
  D.~T.~Son and N.~Yamamoto,
  Phys.\ Rev.\ D {\bf 87}, no. 8, 085016 (2013)
  doi:10.1103/PhysRevD.87.085016
  [arXiv:1210.8158 [hep-th]].
 
\bibitem{Chen:2013iga}
  J.~W.~Chen, J.~y.~Pang, S.~Pu and Q.~Wang,
  Phys.\ Rev.\ D {\bf 89}, no. 9, 094003 (2014)
  doi:10.1103/PhysRevD.89.094003
  [arXiv:1312.2032 [hep-th]].
 
\bibitem{Chen:2014cla} 
  J.~Y.~Chen, D.~T.~Son, M.~A.~Stephanov, H.~U.~Yee and Y.~Yin,
  Phys.\ Rev.\ Lett.\  {\bf 113}, no. 18, 182302 (2014)
  doi:10.1103/PhysRevLett.113.182302
  [arXiv:1404.5963 [hep-th]].
  
  
  \bibitem{Stone:2013sga} 
  M.~Stone and V.~Dwivedi,
  Phys.\ Rev.\ D {\bf 88}, no. 4, 045012 (2013)
  doi:10.1103/PhysRevD.88.045012
  [arXiv:1305.1955 [hep-th]].
  
 \bibitem{Dwivedi:2013dea} 
  V.~Dwivedi and M.~Stone,
  J.\ Phys.\ A {\bf 47}, 025401 (2013)
  doi:10.1088/1751-8113/47/2/025401
  [arXiv:1308.4576 [hep-th]].
  
   
\bibitem{Manuel:2014dza} 
  C.~Manuel and J.~M.~Torres-Rincon,
  Phys.\ Rev.\ D {\bf 90}, no. 7, 076007 (2014)
  doi:10.1103/PhysRevD.90.076007
  [arXiv:1404.6409 [hep-ph]].

\bibitem{Stone:2014fja} 
  M.~Stone, V.~Dwivedi and T.~Zhou,
  Phys.\ Rev.\ D {\bf 91}, no. 2, 025004 (2015)
  doi:10.1103/PhysRevD.91.025004
  [arXiv:1406.0354 [hep-th]].

\bibitem{Manuel:2015zpa} 
  C.~Manuel and J.~M.~Torres-Rincon,
  Phys.\ Rev.\ D {\bf 92}, no. 7, 074018 (2015)
  doi:10.1103/PhysRevD.92.074018
  [arXiv:1501.07608 [hep-ph]].
  
\bibitem{Chen:2015gta} 
  J.~Y.~Chen, D.~T.~Son and M.~A.~Stephanov,
  Phys.\ Rev.\ Lett.\  {\bf 115}, no. 2, 021601 (2015)
  doi:10.1103/PhysRevLett.115.021601
  [arXiv:1502.06966 [hep-th]].

\bibitem{Sun:2016nig} 
  Y.~Sun, C.~M.~Ko and F.~Li,
  Phys.\ Rev.\ C {\bf 94}, no. 4, 045204 (2016)
  doi:10.1103/PhysRevC.94.045204
  [arXiv:1606.05627 [nucl-th]].

\bibitem{Hidaka:2016yjf} 
  Y.~Hidaka, S.~Pu and D.~L.~Yang,
  arXiv:1612.04630 [hep-th].

  
\bibitem{Berry:1984jv} 
  M.~V.~Berry,
  Proc.\ Roy.\ Soc.\ Lond.\ A {\bf 392}, 45 (1984).
  doi:10.1098/rspa.1984.0023

  
\bibitem{Stone:1985av} 
  M.~Stone,
  Phys.\ Rev.\ D {\bf 33}, 1191 (1986).
  doi:10.1103/PhysRevD.33.1191
  
\bibitem{Aitchison:1986qn} 
  I.~J.~R.~Aitchison,
  Acta Phys.\ Polon.\ B {\bf 18}, 207 (1987).

\bibitem{Nelson:1984gu} 
  P.~C.~Nelson and L.~Alvarez-Gaume,
  Commun.\ Math.\ Phys.\  {\bf 99}, 103 (1985).
  doi:10.1007/BF01466595
  
\bibitem{Shapere:1989kp} 
  A.~D.~Shapere and F.~Wilczek,
  Adv.\ Ser.\ Math.\ Phys.\  {\bf 5}, 1 (1989).
 


\bibitem{Deguchi:2005pc} 
  S.~Deguchi and K.~Fujikawa,
  Phys.\ Rev.\ A {\bf 72}, 012111 (2005)
  doi:10.1103/PhysRevA.72.012111
  [hep-th/0501166].
  
\bibitem{Fujikawa:2005tv} 
  K.~Fujikawa,
  Phys.\ Rev.\ D {\bf 72}, 025009 (2005)
  doi:10.1103/PhysRevD.72.025009
  [hep-th/0505087].
  
\bibitem{Fujikawa:2005cn} 
  K.~Fujikawa,
  Phys.\ Rev.\ D {\bf 73}, 025017 (2006)
  doi:10.1103/PhysRevD.73.025017
  [hep-th/0511142].
  
\bibitem{Wess:1971yu} 
  J.~Wess and B.~Zumino,
  Phys.\ Lett.\ B {\bf 37}, 95 (1971).
  doi:10.1016/0370-2693(71)90582-X
  
\bibitem{Witten:1983tw} 
  E.~Witten,
  Nucl.\ Phys.\ B {\bf 223}, 422 (1983).
  doi:10.1016/0550-3213(83)90063-9
  


  
\bibitem{Feynman:1950ir} 
  R.~P.~Feynman,
  Phys.\ Rev.\  {\bf 80}, 440 (1950).
  doi:10.1103/PhysRev.80.440

    
\bibitem{Schwinger:1951nm} 
  J.~S.~Schwinger,
  Phys.\ Rev.\  {\bf 82}, 664 (1951).
  doi:10.1103/PhysRev.82.664
  
\bibitem{Strassler:1992zr} 
  M.~J.~Strassler,
  Nucl.\ Phys.\ B {\bf 385}, 145 (1992)
  doi:10.1016/0550-3213(92)90098-V
  [hep-ph/9205205].


\bibitem{Mondragon:1995va} 
  M.~Mondragon, L.~Nellen, M.~G.~Schmidt and C.~Schubert,
  Phys.\ Lett.\ B {\bf 351}, 200 (1995)
  doi:10.1016/0370-2693(95)00337-K
  [hep-th/9502125].

\bibitem{Mondragon:1995ab} 
  M.~Mondragon, L.~Nellen, M.~G.~Schmidt and C.~Schubert,
  Phys.\ Lett.\ B {\bf 366}, 212 (1996)
  doi:10.1016/0370-2693(95)01392-X
  [hep-th/9510036].
  
  
\bibitem{JalilianMarian:1999xt} 
  J.~Jalilian-Marian, S.~Jeon, R.~Venugopalan and J.~Wirstam,
  Phys.\ Rev.\ D {\bf 62}, 045020 (2000)
  doi:10.1103/PhysRevD.62.045020
  [hep-ph/9910299].
  
  
\bibitem{Schubert:2001he} 
  C.~Schubert,
  Phys.\ Rept.\  {\bf 355}, 73 (2001)
  doi:10.1016/S0370-1573(01)00013-8
  [hep-th/0101036].

\bibitem{Hernandez:2008db} 
  A.~Hernandez, T.~Konstandin and M.~G.~Schmidt,
  Nucl.\ Phys.\ B {\bf 812}, 290 (2009)
  doi:10.1016/j.nuclphysb.2008.12.021
  [arXiv:0810.4092 [hep-ph]].
  
\bibitem{Bastianelli:2006rx} 
  F.~Bastianelli and P.~van Nieuwenhuizen,
  'Path integrals and anomalies in curved space'
    Cambridge University Press (2006)
  
\bibitem{Corradini:2015tik} 
  O.~Corradini and C.~Schubert,
  arXiv:1512.08694 [hep-th].
  

  
    
\bibitem{Polyakov:1987ez} 
  A.~M.~Polyakov,
  Contemp.\ Concepts Phys.\  {\bf 3}, 1 (1987).

\bibitem{Bern:1991aq} 
  Z.~Bern and D.~A.~Kosower,
  Nucl.\ Phys.\ B {\bf 379}, 451 (1992).
  doi:10.1016/0550-3213(92)90134-W
  
\bibitem{Bern:1994zx} 
  Z.~Bern, L.~J.~Dixon, D.~C.~Dunbar and D.~A.~Kosower,
  Nucl.\ Phys.\ B {\bf 425}, 217 (1994)
  doi:10.1016/0550-3213(94)90179-1
  [hep-ph/9403226].
  
\bibitem{AlvarezGaume:1983ig} 
  L.~Alvarez-Gaume and E.~Witten,
  Nucl.\ Phys.\ B {\bf 234}, 269 (1984).
  doi:10.1016/0550-3213(84)90066-X
 
\bibitem{AlvarezGaume:1983at} 
  L.~Alvarez-Gaume,
  Commun.\ Math.\ Phys.\  {\bf 90}, 161 (1983).
  doi:10.1007/BF01205500
  
\bibitem{Fujikawa:1979ay} 
  K.~Fujikawa,
  Phys.\ Rev.\ Lett.\  {\bf 42}, 1195 (1979).
  doi:10.1103/PhysRevLett.42.1195
  
\bibitem{Fujikawa:1980eg} 
  K.~Fujikawa,
  Phys.\ Rev.\ D {\bf 21}, 2848 (1980)
  Erratum: [Phys.\ Rev.\ D {\bf 22}, 1499 (1980)].
  doi:10.1103/PhysRevD.21.2848, 10.1103/PhysRevD.22.1499

\bibitem{D'Hoker:1995ax} 
  E.~D'Hoker and D.~G.~Gagn\'e,
  Nucl.\ Phys.\ B {\bf 467}, 272 (1996)
  doi:10.1016/0550-3213(96)00125-3
  [hep-th/9508131].
  
\bibitem{D'Hoker:1995bj} 
  E.~D'Hoker and D.~G.~Gagn\'e,
  Nucl.\ Phys.\ B {\bf 467}, 297 (1996)
  doi:10.1016/0550-3213(96)00126-5
  [hep-th/9512080].
  
\bibitem{Bargmann:1959gz} 
  V.~Bargmann, L.~Michel and V.~L.~Telegdi,
  Phys.\ Rev.\ Lett.\  {\bf 2}, 435 (1959).
  doi:10.1103/PhysRevLett.2.435
  
\bibitem{Wong:1970fu} 
  S.~K.~Wong,
  Nuovo Cim.\ A {\bf 65}, 689 (1970).
  doi:10.1007/BF02892134


\bibitem{Berezin:1976eg} 
  F.~A.~Berezin and M.~S.~Marinov,
  Annals Phys.\  {\bf 104}, 336 (1977).
  doi:10.1016/0003-4916(77)90335-9
  
\bibitem{Balachandran:1976ya} 
  A.~P.~Balachandran, P.~Salomonson, B.~S.~Skagerstam and J.~O.~Winnberg,
  Phys.\ Rev.\ D {\bf 15}, 2308 (1977).
  doi:10.1103/PhysRevD.15.2308
   
\bibitem{Balachandran:1977ub} 
  A.~P.~Balachandran, S.~Borchardt and A.~Stern,
  Phys.\ Rev.\ D {\bf 17}, 3247 (1978).
  doi:10.1103/PhysRevD.17.3247
  
\bibitem{Barducci:1982yw} 
  A.~Barducci,
  Phys.\ Lett.\ B {\bf 118}, 112 (1982).
  doi:10.1016/0370-2693(82)90611-6

\bibitem{Brink:1976uf} 
  L.~Brink, P.~Di Vecchia and P.~S.~Howe,
  Nucl.\ Phys.\ B {\bf 118}, 76 (1977).
  doi:10.1016/0550-3213(77)90364-9
  
\bibitem{Mathur:1993tp} 
  S.~D.~Mathur,
  hep-th/9306090.
 
\bibitem{Bodeker:1998hm} 
  D.~Bodeker,
  Phys.\ Lett.\ B {\bf 426}, 351 (1998)
  doi:10.1016/S0370-2693(98)00279-2
  [hep-ph/9801430].
  
\bibitem{Bodeker:1999ey} 
  D.~B\"{o}deker,
  Nucl.\ Phys.\ B {\bf 559}, 502 (1999).
  


  \bibitem{Litim:1999id} 
  D.~F.~Litim and C.~Manuel,
  Nucl.\ Phys.\ B {\bf 562}, 237 (1999).
  
\bibitem{follow-up}
N.~Mueller, R. Venugopalan and Y.~Yin, in progress. 

\bibitem{Mehta:1986mi} 
  M.~R.~Mehta,
  Phys.\ Rev.\ Lett.\  {\bf 65}, 1983 (1990)
  Erratum: [Phys.\ Rev.\ Lett.\  {\bf 66}, 522 (1991)].
  doi:10.1103/PhysRevLett.65.1983


  
\bibitem{Ohnuki:1978jv} 
  Y.~Ohnuki and T.~Kashiwa,
  Prog.\ Theor.\ Phys.\  {\bf 60}, 548 (1978).
  doi:10.1143/PTP.60.548
  
\bibitem{FiorenzoBookNew}
  F.~Bastianelli and C. Schubert, book in preparation (Cambridge University Press) and F.~Bastianelli, {\it Constrained hamiltonian systems and
relativistic particles}, lecture notes INFN Bologna 2012.

\bibitem{Barducci:1976xq} 
  A.~Barducci, R.~Casalbuoni and L.~Lusanna,
  Nucl.\ Phys.\ B {\bf 124}, 93 (1977).


\bibitem{Sakurai}J. J. Sakurai, {\it Advanced Quantum Mechanics}, Addison-Wesley Publishers (1967). 


\bibitem{Bardeen:1984pm} 
  W.~A.~Bardeen and B.~Zumino,
  Nucl.\ Phys.\ B {\bf 244}, 421 (1984).
  doi:10.1016/0550-3213(84)90322-5
  
\bibitem{Bardeen:1969md} 
  W.~A.~Bardeen,
  Phys.\ Rev.\  {\bf 184}, 1848 (1969).
  doi:10.1103/PhysRev.184.1848
  
\bibitem{Dunne:2005sx} 
  G.~V.~Dunne and C.~Schubert,
  Phys.\ Rev.\ D {\bf 72}, 105004 (2005)
  doi:10.1103/PhysRevD.72.105004
  [hep-th/0507174].
  
\bibitem{Dunne:2006st} 
  G.~V.~Dunne, Q.~h.~Wang, H.~Gies and C.~Schubert,
  Phys.\ Rev.\ D {\bf 73}, 065028 (2006)
  doi:10.1103/PhysRevD.73.065028
  [hep-th/0602176].


\bibitem{Simon:1983mh} 
  B.~Simon,
  Phys.\ Rev.\ Lett.\  {\bf 51}, 2167 (1983).
  doi:10.1103/PhysRevLett.51.2167
  
\bibitem{Akamatsu:2014yza} 
  Y.~Akamatsu and N.~Yamamoto,
  Phys.\ Rev.\ D {\bf 90}, no. 12, 125031 (2014)
  doi:10.1103/PhysRevD.90.125031
  [arXiv:1402.4174 [hep-th]].
  
  
  
\bibitem{Akamatsu:2015kau} 
  Y.~Akamatsu, A.~Rothkopf and N.~Yamamoto,
  JHEP {\bf 1603}, 210 (2016)
  doi:10.1007/JHEP03(2016)210
  [arXiv:1512.02374 [hep-ph]].

 
 \bibitem{Charbonneau:2009ax} 
  J.~Charbonneau and A.~Zhitnitsky,
  JCAP {\bf 1008}, 010 (2010).

\bibitem{Akamatsu:2013pjd} 
  Y.~Akamatsu and N.~Yamamoto,
  Phys.\ Rev.\ Lett.\  {\bf 111}, 052002 (2013).

\bibitem{Dvornikov:2016gdo} 
  M.~Dvornikov and V.~B.~Semikoz,
  arXiv:1603.07946 [astro-ph.CO].

\bibitem{Kaplan:2016drz} 
  D.~B.~Kaplan, S.~Reddy and S.~Sen,
  arXiv:1612.00032 [hep-ph].

  
\bibitem{Affleck:1981bma} 
  I.~K.~Affleck, O.~Alvarez and N.~S.~Manton,
  Nucl.\ Phys.\ B {\bf 197}, 509 (1982).
  doi:10.1016/0550-3213(82)90455-2
  
\bibitem{Kovchegov:2016weo} 
  Y.~V.~Kovchegov, D.~Pitonyak and M.~D.~Sievert,
  arXiv:1610.06188 [hep-ph].

\bibitem{McLerran:1993ni} 
  L.~D.~McLerran and R.~Venugopalan,
  Phys.\ Rev.\ D {\bf 49}, 2233 (1994)
  doi:10.1103/PhysRevD.49.2233
  [hep-ph/9309289].

\bibitem{McLerran:1993ka} 
  L.~D.~McLerran and R.~Venugopalan,
  Phys.\ Rev.\ D {\bf 49}, 3352 (1994)
  doi:10.1103/PhysRevD.49.3352
  [hep-ph/9311205].

\bibitem{JalilianMarian:2000ad} 
  J.~Jalilian-Marian, S.~Jeon and R.~Venugopalan,
  Phys.\ Rev.\ D {\bf 63}, 036004 (2001)
  doi:10.1103/PhysRevD.63.036004
  [hep-ph/0003070].
  
\bibitem{Kuraev:1977fs} 
  E.~A.~Kuraev, L.~N.~Lipatov and V.~S.~Fadin,
  Sov.\ Phys.\ JETP {\bf 45}, 199 (1977)
  [Zh.\ Eksp.\ Teor.\ Fiz.\  {\bf 72}, 377 (1977)].
  
\bibitem{Balitsky:1978ic} 
  I.~I.~Balitsky and L.~N.~Lipatov,
  Sov.\ J.\ Nucl.\ Phys.\  {\bf 28}, 822 (1978)
  [Yad.\ Fiz.\  {\bf 28}, 1597 (1978)].



 
  
\end{thebibliography}
\end{document}